\documentclass[acmsmall]{acmart}
\usepackage{tabularx}
\usepackage{graphicx}
\usepackage{longtable}
\usepackage{tcolorbox}
\usepackage{hyperref}
\usepackage[utf8]{inputenc}
\DeclareUnicodeCharacter{2060}{}

\AtBeginDocument{%
  }

\setcopyright{acmlicensed}
\copyrightyear{2025}
\acmYear{2025}
\acmDOI{XXXXXXX.XXXXXXX}

\begin{document}

\title{MITRE ATT\&CK Applications in Cybersecurity and The Way Forward}


\author{Yuning Jiang}
\email{yuning\_j@nus.edu.sg}
\orcid{0000-0003-4791-8452}
\affiliation{%
  \institution{National University of Singapore}
  \country{Singapore}
}

\author{Qiaoran Meng}
\email{qiaoran@nus.edu.sg}
\affiliation{%
  \institution{National University of Singapore}
  \country{Singapore}}

\author{Feiyang Shang}
\email{feiyang.shang@u.nus.edu}
\affiliation{%
  \institution{National University of Singapore}
  \country{Singapore}}
  
\author{Nay Oo}
\email{nay.oo@ncs.com.sg}
\affiliation{%
 \institution{NCS Cyber Special Ops R\&D}
  \country{Singapore}}

\author{Le Thi Hong Minh}
\affiliation{%
  \institution{National University of Singapore}
  \country{Singapore}}
\email{e1297762@u.nus.edu}

\author{Hoon Wei Lim}
\email{hoonwei.lim@ncs.com.sg}
\affiliation{%
 \institution{NCS Cyber Special Ops R\&D}
  \country{Singapore}}

\author{Biplab Sikdar}
\email{bsikdar@nus.edu.sg}
\affiliation{%
  \institution{National University of Singapore}
  \country{Singapore}}

\renewcommand{\shortauthors}{Jiang et al.}

\begin{abstract}

The MITRE ATT\&CK framework is a widely adopted tool for enhancing cybersecurity, supporting threat intelligence, incident response, attack modeling, and vulnerability prioritization. This paper synthesizes research on its application across these domains by analyzing 417 peer-reviewed publications. We identify commonly used adversarial tactics, techniques, and procedures (TTPs) and explore the integration of natural language processing (NLP) and machine learning (ML) with ATT\&CK to improve threat detection and response. Additionally, we examine ATT\&CK’s integration with other frameworks, such as the Cyber Kill Chain, NIST guidelines, and STRIDE, highlighting its versatility. The paper also evaluates ATT\&CK’s effectiveness, validation methods, and challenges in sectors like industrial control systems (ICS) and healthcare. We conclude by discussing current limitations and proposing future research directions to enhance ATT\&CK’s application in dynamic cybersecurity environments.

\end{abstract}




\maketitle

\section{Introduction}

MITRE ATT\&CK (or Adversarial Tactics, Techniques, and Common Knowledge) is a curated knowledge base that details cyber adversary behavior across various domains, including enterprise, mobile, and industrial control systems (ICS) \cite{strom2018mitre, son2023introduction}. Organized into matrices, ATT\&CK outlines specific adversarial tactics (high-level objectives) and techniques (specific actions) that describe how an attacker seeks to compromise or persist in a system. 

MITRE ATT\&CK has become a foundational tool in cybersecurity practices. By enabling the mapping of adversarial tactics, techniques, and procedures (TTPs), ATT\&CK enhances the systematic detection and analysis of cyber threats, particularly advanced persistent threats (APTs) \cite{golushko2022adversary}. The framework has been widely applied across industries like healthcare, finance, and critical infrastructure, where its TTPs significantly contribute to enhanced threat detection and mitigation through tools such as AI, ML, and endpoint detection and response (EDR) systems \cite{roy2023sok, oosthoek2019sok, al2024mitre}.

However, despite its widespread use, there remains a need to explore ATT\&CK’s integration within broader cybersecurity methodologies and to address the technical challenges associated with its application. This paper aims to systematically review the academic literature on MITRE ATT\&CK, consolidating insights into its applications, technical integration, and the limitations researchers have encountered. The review will also identify opportunities for future research, proposing directions for refining and extending the framework to address existing gaps.

The research questions specifically addressed by this study are as follows:
\begin{itemize}
    \item[RQ1:] Why is it important to adopt the MITRE ATT\&CK framework, what are its primary uses, and how effective is its application across different cybersecurity domains?

    \item[RQ2:] What are the technical methods, tools, and data sources used in studies applying the MITRE ATT\&CK framework, and how are the results validated?

    \item[RQ3:] What are the common challenges, limitations, and future directions identified in the literature regarding the use of MITRE ATT\&CK?

\end{itemize}

This paper analyzes the application of the MITRE ATT\&CK framework across cybersecurity domains, such as threat intelligence, incident response, attack modeling, and vulnerability prioritization. Based on a review of 417 articles, we identify frequently employed TTPs and their use in industries like enterprise systems, manufacturing, and healthcare. We also highlight the integration of natural language processing (NLP) and machine learning (ML) with ATT\&CK, improving adversarial behavior detection and threat response. We examine ATT\&CK's integration with other frameworks, such as the Cyber Kill Chain, NIST guidelines, STRIDE, and CIS Controls, emphasizing its adaptability in fostering a comprehensive cybersecurity approach. Additionally, we evaluate various validation methods, including case studies and empirical experiments, to assess ATT\&CK’s effectiveness. This paper provides insights for researchers and practitioners seeking to enhance cybersecurity strategies while addressing ATT\&CK's challenges and limitations.

This paper is structured to guide readers from foundational concepts to actionable insights, identifying research gaps and future directions. Section \ref{sec:Preliminaries} introduces key concepts, while Section \ref{sec:Methodology} outlines the systematic review process. Core findings are detailed in Sections \ref{sec:ResearchFocus}–\ref{sec:Validation}, covering research focuses, data sources, tools, ATT\&CK integrations, and validation methods. Section \ref{sec:Discussion} synthesizes these findings, highlighting challenges and gaps, and Section \ref{sec:Conclusion} summarizes insights and future directions. A conceptual diagram is included to visually illustrate the organization and relationships between sections, enhancing clarity and navigation.

\section{Preliminaries} 
\label{sec:Preliminaries}

This section provides an overview of foundational concepts, establishing key background and terminology for the study.

\subsection{MITRE ATT\&CK}

Developed in 2013 as part of MITRE's Fort Meade Experiment (FMX), MITRE ATT\&CK \cite{strom2018mitre} is a research initiative aimed at enhancing forensic analysis of cyberattacks by emulating adversarial and defensive tactics. The framework serves as a comprehensive knowledge base of real-world adversarial techniques, focusing on how attackers interact with systems rather than the tools or malware they use. It is structured around four core components: tactics (adversarial goals), techniques (methods to achieve goals), sub-techniques (detailed methods breaking down techniques), and documented usage (procedures and metadata).

MITRE ATT\&CK is organized into three technology domains representing the environments where adversaries operate: Enterprise (traditional networks and cloud technologies), Mobile (mobile devices), and ICS (Industrial Control Systems). Each domain features distinct tactics and techniques, with techniques further detailed into sub-techniques, describing specific adversarial behaviors. Complementing ATT\&CK, the MITRE FiGHT framework addresses supply chain threats, mapping 5G-relevant techniques to attack tactics.

\subsection{Related Works}

Al-Sada Bader et al. \cite{al2024mitre} categorize 50 research contributions based on various use cases, including behavioral analytics, red teaming, defensive gap assessment, and CTI enrichment. This categorization highlights different applications, methodologies, and data sources in the implementation of ATT\&CK, with a focus on adversarial behavior modeling, automated threat detection, and framework enhancements. The categorization draws from a white paper \cite{strom2018mitre} discussing the design philosophy, structure, and use cases of MITRE ATT\&CK, and a similar white paper \cite{alexander2020mitre} on the development and application of the ATT\&CK for ICS framework.

Roy Shanto et al. \cite{roy2023sok} provide a taxonomic classification of ATT\&CK-related research, identifying use cases, application domains, and methodologies. Key areas include CTI, intrusion detection, offensive security, cyber risk assessment, professional training, and threat-driven approaches. This study also identifies a gap between academic and industry use of ATT\&CK, with industry focusing on practical implementations and tools, while academia explores theoretical models.

Several studies focus on specific applications of ATT\&CK. Kris Oosthoek and Christian Doerr \cite{oosthoek2019sok} analyze 951 Windows malware families from the Malpedia repository, mapping post-compromise malware techniques to ATT\&CK and examining techniques observed in a controlled sandbox environment. Joshua Bolton et al. \cite{bolton2023overview} explore the use of knowledge graphs (KG) in cybersecurity, particularly in conjunction with MITRE ATT\&CK.

Given the rapid evolution of cyber threats and the increasing complexity of defensive strategies, understanding the MITRE ATT\&CK framework's role across cybersecurity domains is crucial. Despite its widespread adoption, there is limited systematic analysis of ATT\&CK’s application across sectors, methodologies, and advanced technologies. To address this, we conduct a comprehensive literature review analyzing ATT\&CK’s application in threat intelligence, incident response, attack modeling, and vulnerability prioritization.

\section{Methodology for Systematization}
\label{sec:Methodology}

This section presents the methodology of our systematization in a step-by-step manner, and clearly defines the criteria for literature selection, databases, and search keywords used, as well as inclusion and exclusion criteria.

\subsection{Databases and Search Keywords}

A comprehensive literature search was conducted on May 31, 2024 in three digital libraries, namely IEEE Xplore, Scopus, and the ACM Digital Library. The initial search yielded 738 results from IEEE Xplore, which, after filtering out non-research content such as books and magazines, was reduced to 563. Similarly, the search in the ACM Digital Library resulted in 23 entries after removing non-research materials. Scopus produced 922 initial results, which were reduced to 847 after excluding non-research content. After removing duplicates across these databases, a total of 1,168 unique papers remained in the pool for further evaluation.

\subsection{Inclusion and Exclusion Criteria}

We assessed the relevance of papers through a multi-phase screening process. In the initial phase, we reviewed titles, abstracts, and keywords for alignment with key MITRE ATT\&CK concepts, including threat modeling, intrusion detection, incident management, threat intelligence, attack modeling, and vulnerability analysis. If the abstract alone did not provide sufficient information, the full text was reviewed. Papers were included if they applied, extended, or compared MITRE ATT\&CK with other frameworks, methodologies, or tools. Any ambiguity regarding inclusion was resolved through discussion among the research team. Exclusion criteria include:

\begin{itemize}
    \item \textbf{Relevance:} Papers not directly addressing the research questions.
    \item \textbf{MITRE ATT\&CK Elaboration:} Papers that only mentioned MITRE ATT\&CK without detailed elaboration.
    \item \textbf{Language:} Non-English papers were excluded.
    \item \textbf{Peer Review:} Only peer-reviewed papers were considered.
    \item \textbf{Length:} Papers shorter than four pages and lacking sufficient detail were excluded.
\end{itemize}

The screening process occurred in two rounds. In the first, papers were filtered based on abstracts and keywords. A total of 419 papers were excluded. This phase was conducted by two cybersecurity experts, with two additional researchers independently verifying the final selection. After filtering, 396 papers remained for in-depth review.

To ensure comprehensive coverage, both backward and forward snowballing techniques were applied. In forward snowballing, we tracked citations of selected papers using Google Scholar, ensuring the inclusion of recent studies. In backward snowballing, we reviewed references and citations within those references until no additional relevant papers were found. Each paper identified through snowballing was re-evaluated against the inclusion and exclusion criteria. This process led to the addition of 21 papers, bringing the total to 417 papers \footnote{https://github.com/LeThiHongMinh/MITRE-review-papers} included in the review.

\subsection{Data Extraction}

We employed a structured data extraction process to systematically gather relevant information from the selected studies. Using a predefined template, we extracted bibliographic details such as authors, title, publication year, source, document type, abstract, and keywords. This ensured consistency and facilitated subsequent analysis. In addition, we manually categorized each paper based on its methodologies, theoretical frameworks, evaluation techniques, and real-world applications. Key features extracted included:

\begin{itemize}
    \item The primary goals and questions, particularly how MITRE ATT\&CK was employed.
    \item The domain or specific focus (e.g., threat intelligence) for categorizing application areas.
    \item The methodologies used in each study (e.g., empirical experiments, case studies, simulations, surveys) was detailed, providing insight into how MITRE ATT\&CK was operationalized.
    \item ATT\&CK tactics, techniques, and integration with other frameworks (e.g., Cyber Kill Chain).
    \item The tools, techniques, data sources, and approaches used in each paper were cataloged.
    \item The main insights and the effectiveness of MITRE ATT\&CK in cybersecurity applications. 
    \item Evaluation approaches used (e.g., experimental validation, real-world case validation).
    \item Issues with MITRE ATT\&CK’s application, such as integration difficulties.
    \item Suggested enhancements and areas for further research.
\end{itemize}

This extraction was carried out collaboratively by two researchers and validated by two additional researchers to minimize misinterpretation or coding bias.

\subsection{Analysis Strategy}

Our analysis strategy synthesized the extracted data to identify broader themes and trends across the literature. We first summarized key features of each study into thematic categories (e.g., research focus areas, methodologies, technical tools, and evaluation methods). These summaries were reviewed to uncover recurring patterns and relationships.

We also mapped the relationships among predominant methodologies, design features, and frameworks used in the literature. This process facilitated the exploration of potential synthesis and cross-pollination of ideas across studies, providing a comprehensive overview of how MITRE ATT\&CK has been applied, extended, or integrated with other frameworks in the field.

\section{Applications of MITRE ATT\&CK} \label{sec:ResearchFocus}

This section synthesizes key focus areas and industry-specific applications in MITRE ATT\&CK research.

\subsection{Research Focus Areas} \label{sec:FocusAreas}

The MITRE ATT\&CK framework is widely applied across cybersecurity research domains, with ten key research areas identified in the reviewed papers, as summarized in Table~\ref{tab:focus}. These ten key areas were built on top of and extended from the research areas identified in \cite{roy2023sok, alexander2020mitre}. Each paper is categorized into one to three primary areas, with uncategorized papers grouped under \textit{Others}. CTI and Threat Hunting dominate ATT\&CK-related research, accounting for 30.2\% (126 out of 417) and 17.5\% (73 out of 417) of the papers, respectively.

\begin{table}[h]
\centering
\footnotesize
\caption{Research Focus Areas of MITRE ATT\&CK Framework.}
\label{tab:focus}
\resizebox{\textwidth}{!}{ 
\begin{tabularx}{\textwidth}{|X|l|l|}
    \hline
    \textbf{Research Focus Area} & Frequency & \textbf{Example Papers} \\ 
    \hline
    \textbf{Cyber Threat Intelligence}: Evidence-based knowledge about cyber threats, including indicators of compromise, adversary tactics, and threat actor profiles. & 126 & \cite{li2024automated, huang2024mitretrieval, rani2023ttphunter, irshad2023cyber}\\ \hline
    \textbf{Threat Hunting}: The proactive approach of identifying and mitigating potential cyber threats that may have bypassed security measures in the system. & 73 & \cite{sharma2023radar, kang2022actdetector, takey2022real, kuwano2022att} \\ \hline
    \textbf{Threat Modeling}: A systematic framework for identifying potential system vulnerabilities, attack vectors, and their potential impacts. & 48 & \cite{xiong2022cyber, hotellier2024standard, rao2023threat, pell2021towards} \\ \hline
    \textbf{Attack Simulation}: The process of simulating adversary attacks in cyber systems for penetration testing, risk assessment, and other security objectives. & 44 & \cite{mohamed2022study, takahashi2020aptgen, choi2021probabilistic, toker2021mitre} \\ \hline
    \textbf{Risk Assessment}: Evaluating cybersecurity posture, compliance, and risks from vulnerabilities and threats. & 43 & \cite{falco2018master, georgiadou2021assessing, manocha2021security, oruc2022assessing} \\ \hline
    \textbf{Incident Response}: A structured methodology for detecting, containing, and recovering from security breaches while minimizing impact. & 37 & \cite{leite2022actionable, boakye2023implementation, nisioti2021data, outkin2022defender} \\ \hline
    \textbf{APT Detection}: The process of identifying sophisticated, long-term cyber attacks conducted by well-resourced adversaries using stealth techniques. & 27 & \cite{jabar2022mobile, mahmoud2023apthunter, soliman2023rank, park2023performance} \\ \hline
    \textbf{Intrusion Detection System}: A security mechanism that monitors activities for malicious behavior and policy violations using signature or anomaly methods. & 25 & \cite{wang2022ai, arreche2024xai, hermawan2021development, landauer2022maintainable} \\ \hline
    \textbf{Malware Detection}: The identification of malicious software, including ransomware, designed to compromise system security or functionality. & 22 & \cite{gulatas2023malware, chimmanee2024digital, ahn2023exploring, villanueva2022detecting} \\ \hline
    Others & 14 & \cite{rostami2020machine, karch2022crosstest} \\ \hline
    \textbf{Vulnerability Management}: The systematic process of identifying, assessing, prioritizing, and addressing security vulnerabilities in systems and software. & 11 & \cite{hacks2022multi, touloumis2021vulnerabilities, preethi2023leveraging, devaux2021automation} \\ \hline
\end{tabularx}}  
\end{table}

\subsubsection{Cyber Threat Intelligence (CTI)} \label{sec:CTI}

Research leveraging the MITRE ATT\&CK framework focuses on extracting CTI information from raw data through the application of NLP and ML techniques. Studies such as \cite{li2024automated, huang2024mitretrieval, kim2022comparative, liu2022threat, rani2023ttphunter} have proposed frameworks for extracting TTPs from unstructured reports using LLMs and transformer-based architectures like BERT \cite{li2024automated, huang2024mitretrieval, rani2023ttphunter}. Quantitative evaluations across large, TTP-labeled datasets indicate comprehensive coverage of MITRE ATT\&CK tactics and techniques in \cite{huang2024mitretrieval, kim2022comparative, liu2022threat}, whereas \cite{li2024automated, rani2023ttphunter} focus on subsets of techniques. For instance, \cite{li2024automated} employs DistilBERT-based multi-label classifiers to categorize tactics and techniques, incorporating pre-processing for CTI text denoising and post-processing for correction. Similarly, \cite{huang2024mitretrieval} integrates ontology knowledge with sentence-level BERT models, combining classification outcomes using a voting algorithm to address the lack of labeled datasets. \textsc{TTPHunter} by \cite{rani2023ttphunter} refines BERT embeddings for extracting TTPs from APT reports. In addition to LLMs, innovative approaches such as \cite{liu2022threat} introduce an Attention-based Transformer Hierarchical Recurrent Neural Network, while \cite{kim2022comparative} enhances small, imbalanced datasets with data augmentation techniques for ML classifiers.

Broader CTI applications expand beyond TTPs to include malware profiles, attack tools, and other features \cite{tang2023attack, irshad2023cyber, alam2023looking}. For example, \cite{irshad2023cyber} proposed \textsc{Attack2Vec}, a domain-specific text embedding model extracting CTA features from CTI reports for classification. \cite{tang2023attack} integrates ATT\&CK-based networks with BERT for semantic analysis, categorizing malware and other CTAs. Similarly, \textsc{LADDER} by \cite{alam2023looking} extracts entities and relationships for TTP classification and threat intelligence knowledge graph construction, aiding campaign inference from new CTI sources. This work also introduces an open-source benchmark dataset for malware-related CTI studies.

\subsubsection{Threat Hunting}

MITRE ATT\&CK framework serves as a foundational tool for threat hunting, particularly in network traffic analysis, system log analysis, and attack graph modeling. 

Several studies utilize ATT\&CK for analyzing network traffic to classify malicious activities \cite{sharma2023radar, bagui2023using, kang2022actdetector}. \textsc{RADAR} \cite{sharma2023radar} classifies network logs using an ontology of adversary behavior, combining explainable classifiers with detection rules for TTP classification. However, \textsc{RADAR} covers only 13 techniques across six tactics. Similarly, \textsc{ActDetector} \cite{kang2022actdetector} processes NIDS alerts to detect attack phases aligned with 14 ATT\&CK tactics. Using Doc2Vec embeddings and temporal sequence classifiers, it identifies multi-stage attack activities. Graph-based approaches, such as \cite{bagui2023using}, leverage Neo4j for representing network traffic patterns, focusing on the reconnaissance tactic.

System logs are extensively used to map ATT\&CK techniques for threat detection \cite{rodriguez2023discovering, takey2022real, kuwano2022att, okada2024predicting}. \cite{rodriguez2023discovering} employs process mining to detect adversarial patterns by profiling system event logs into TTPs and mining process models. Similarly, \cite{takey2022real} integrates IoC-based ATT\&CK mapping with ML models to detect multi-stage attacks in real time. In lateral movement detection, \cite{okada2024predicting} extracts ATT\&CK techniques from Windows logs and identifies movement paths through quantitative analysis. A recommendation system proposed in \cite{kuwano2022att} predicts attacker behaviors using KNN-based collaborative filtering, correlating logs with ATT\&CK TTPs.

Attack graphs provide a comprehensive view of multi-step attack scenarios by integrating ATT\&CK tactics. One paper \cite{sadlek2022identification} has combined MITRE ATT\&CK framework and attack graph for attack path identification in multi-step cyber threat scenarios. Similar to \cite{kang2022actdetector}, ATT\&CK tactics are utilized as kill chain stages. The attack graph, constructed based on threat and security property violation categories, contains asset property, attack technique, goal and countermeasure information for effective security investigation.

\subsubsection{Threat Modeling}

MITRE ATT\&CK framework has been extensively applied across diverse sectors for cyber threat modeling, including enterprise IT \cite{xiong2022cyber}, ICS \cite{hotellier2024standard, fujita2023structured, zhang2022threat}, mobile networks \cite{van2023mitre, rao2023threat}, and 5G networks \cite{vanderveen2022threat, santos2022threat, pell2021towards}.

A threat modeling language tailored for enterprise IT is proposed in \cite{xiong2022cyber}, integrating the MITRE ATT\&CK Enterprise Matrix with meta-language framework (MAL). The language defines a entity-relationship model, covering assets and their associations, attack steps represented by ATT\&CK techniques, and defenses. 

In ICS, ATT\&CK-based methodologies focus on enhancing system security \cite{hotellier2024standard} \cite{fujita2023structured} \cite{zhang2022threat}. A specification-based IDS correlating safety standards with attack properties is proposed in \cite{hotellier2024standard}, whereby MITRE ATT\&CK is integrated in process-aware attack modeling as the basis attack framework. \cite{fujita2023structured} proposed a structured model to represent cyber attack scenarios in ICS based on the diamond model. Attack state is defined in the attacker behavioral model part, including attack phrase, capability, location and other properties. Additionally, \cite{zhang2022threat} introduces an ICS-specific security ontology grounded in the Purdue model for effective asset-based threat reasoning.

Threat modeling research has extensively applied the MITRE ATT\&CK framework to both mobile \cite{van2023mitre, rao2023threat} and 5G \cite{vanderveen2022threat, santos2022threat, pell2021towards} networks, addressing shared and unique challenges within these domains. Mobile network threat modeling focuses on categorizing known attacks against cellular communications and extending ATT\&CK’s Mobile Matrix. For example, \cite{van2023mitre} extends ATT\&CK’s mobile matrix into more threat behaviors in 5G by integrating it with the CONCORDIA Mobile Threat Modelling Framework (CMTMF), addressing gaps in ATT\&CK’s mobile threat coverage. The techniques in mobile matrix is mapped to techniques in the enterprise matrix to form a extended ATT\&CK mobile network matrix, and subsequently integrated into other mobile frameworks. Similarly, \cite{rao2023threat} categorizes cellular attack TTPs using ATT\&CK tactics and maps them to the Bhadra framework for structured analysis. In the context of 5G networks, \cite{vanderveen2022threat} introduces \textsc{FiGHT}, a 5G-specific threat modeling framework, which extracts granular attack behaviors from existing publications and maps them to adversarial techniques using ATT\&CK. Similarly, \cite{santos2022threat} constructs the mobile threat modeling framework named \textsc{CMTMF} to cover a broader range of 5G-specific tactics and techniques, incorporating emerging adversarial strategies into a unified framework. \cite{pell2021towards} further enriches 5G modeling by mapping adversarial TTPs to core network components, leveraging historical attacks, threat assessments, and industry reports to provide a comprehensive view of 5G vulnerabilities.

Beyond IT and telecommunications, MITRE ATT\&CK is applied in finance \cite{alevizos2023cyber}, transport \cite{yekta2023vatt}, wastewater systems \cite{davis2024cyber}, maritime security \cite{yousaf2024sinking} and several other domains, showcasing its versatility across critical infrastructure domains (detailed in Section \ref{sec:ApplicationDomain}).

\subsubsection{Attack Simulation}

MITRE ATT\&CK framework is a cornerstone for modeling adversarial TTPs in attack simulation research, leveraging its matrices for various system domains. Researchers have utilized the Enterprise and ICS matrices for simulating attacks in enterprise IT systems and operational technology (OT) environments \cite{ajmal2021offensive, mohamed2022study, takahashi2020aptgen, choi2021probabilistic, toker2021mitre}.

Adversary emulation is a prominent approach in attack simulation \cite{ajmal2021offensive, toker2021mitre}. \cite{ajmal2021offensive} constructs attack emulation plans using CTI-derived TTPs to simulate attack flows, enabling proactive threat hunting and hypothesis validation. Similarly, \cite{toker2021mitre} applies ATT\&CK ICS techniques to simulate multi-step attack scenarios on water management systems. Detection methods, including SVM-based classifiers and Wazuh host-based intrusion detection system (HIDS), are evaluated to visualize attack propagation on an ELK dashboard.

Attack simulations also support the generation of datasets for incident response and methodology evaluation \cite{takahashi2020aptgen, choi2021probabilistic}. \cite{takahashi2020aptgen} introduces \textsc{APTGen}, a top-down framework leveraging ATT\&CK Enterprise techniques to generate targeted attack datasets. The process includes attack sequence generation from cybersecurity reports and selecting specific attack tools based on environmental contexts. Similarly, \cite{choi2021probabilistic} employs a hidden semi-Markov model to generate probabilistic attack sequences based on ATT\&CK ICS TTPs, tested on the HAI testbed using Metasploit.

Platforms like MITRE Caldera \cite{chetwyn2024modelling, lee2023codex, noel2023graph, gjerstad2022lademu, mohamed2022study} automate attack simulations, offering a flexible tool for executing TTP sequences. \cite{mohamed2022study} demonstrates Caldera’s effectiveness in simulating APT attacks that bypass Windows 10 security measures using GoLang agents, providing actionable recommendations for mitigating identified vulnerabilities. 

\subsubsection{Risk Assessment}

MITRE ATT\&CK framework is widely adopted for its clear categorization of cyber threat actors, with both the enterprise matrix \cite{manocha2021security, derbyshire2021talking, ahmed2022mitre} and ICS ATT\&CK matrix \cite{amro2023cyber, adaros2021indicators, brenner2023better, yockey2023cyber, yoon2023vulnerability} integrated into risk assessment methodologies. Risk assessment is often combined with other research areas, such as vulnerability management \cite{yoon2023vulnerability, muthia2023risk} and threat modeling \cite{naik2022evaluation, mondal2023security, lanni2024boosting}, for a more comprehensive security evaluation.

In risk assessment, ATT\&CK Enterprise TTPs are used for attack graph construction to assess the risks of successful attacks \cite{adaros2021indicators}. Several studies \cite{manocha2021security, derbyshire2021talking, ahmed2022mitre} focus on using ATT\&CK Enterprise tactics to calculate system risk ratings and scores. For example, \cite{manocha2021security} computes risk scores based on system security tests and ATT\&CK tactics, while \cite{derbyshire2021talking} proposes a framework for adversary attack cost estimation, developing security controls based on ATT\&CK tactics. Similarly, \cite{ahmed2022mitre} introduces a cyber digital twin, linking past attacks to ATT\&CK tactics and using attack graph analytics to assess and prioritize security control impacts.

The ATT\&CK ICS framework is applied across multiple industries, including maritime \cite{oruc2022assessing, amro2023cyber, amro2023evaluation}, energy \cite{gourisetti2022assessing, mccarty2023cybersecurity, ullah2019cyber}, water \cite{alfageh2023water}, and general ICS systems. ATT\&CK is used as a knowledge base to assess and quantify risks in maritime cyber components \cite{oruc2022assessing} and to enhance current CPS risk evaluation methodologies with semantic knowledge \cite{amro2023evaluation}. In the energy sector, risks are mapped to the ATT\&CK ICS matrix for power systems applications \cite{gourisetti2022assessing}, and multi-step attack scenarios are analyzed for impact in cyber-physical wind energy sites \cite{mccarty2023cybersecurity}. Additionally, \cite{adaros2021indicators} uses the ICS matrix to construct a knowledge base for continuous ICS risk monitoring. Automation of risk assessment in manufacturing systems is explored in \cite{ehrlich2023determining}, where the ATT\&CK ICS framework is integrated with industrial standards to assess the security level of cyber components.

\subsubsection{Incident Response}

MITRE ATT\&CK framework provides detection sources and mitigation measures for attack techniques, making it a critical component in incident response research. ATT\&CK has been applied to the selection and evaluation of security controls \cite{rahman2022investigation, tavolato2024comparing, mohamed2024enhancing}. For example, \cite{rahman2022investigation} examines current security controls, assessing their effectiveness based on mappings between controls and ATT\&CK techniques. Similarly, \cite{mohamed2024enhancing} integrates ATT\&CK into a multi-criteria decision-making methodology to prioritize security controls in defense planning.

ATT\&CK is also leveraged for defense policy evaluation in empirical studies \cite{outkin2022defender, su14031256, boakye2023implementation}. \cite{outkin2022defender} uses a game-theory approach to optimize defense strategies against multi-step attacks, evaluated with ATT\&CK data. \cite{su14031256} focuses on in-network deception technology for defense, evaluating it against ATT\&CK ICS attack profiles. \cite{boakye2023implementation} proposes a comprehensive smart grid attack defense framework, conducting model tests on ATT\&CK ICS-based attacks.

Several studies focus on attack detection and mitigation as part of incident response plans. \cite{rajesh2022analysis} compares threat detection methods and analyzes ATT\&CK-based mitigations. \cite{mundt2023threat} focuses on mitigating data exfiltration, using simulations based on automatically extracted ATT\&CK attack data to generate potential mitigations. Additionally, some incident response works integrate CTI, using ATT\&CK as a CTI source for attack techniques and mitigations \cite{leite2022actionable}.

\subsubsection{Intrusion Detection System}

IDS relies on inbuilt detection rules and external CTI sources to detect suspicious activities in the cyber system, in IDS related research, ATT\&CK framework is widely utilized in AI-based IDS enhancements \cite{wang2022ai, arreche2024xai, arreche2024two, hermawan2021development}. In \cite{wang2022ai}, MITRE ATT\&CK tactics are utilized for attack profiling for Linux system interaction information. Feature extraction and threat detection modules are subsequently developed based on the labelled attacks. \cite{hermawan2021development} designs a threat hunting platform similar to an IDS. MITRE ATT\&CK and diamond model are utilized for incident analysis and detection rule generation. Another use of ATT\&CK lies in the generation and study of IDS datasets \cite{zacharis2023aicef, dehlaghi2023anomaly, landauer2022maintainable, borisenko2022cse}. These datasets are usually generated by IDS from benign network activities and simulated attacks, where ATT\&CK is utilised as the attack generation and data labelling framework. For example, \cite{zacharis2023aicef} aims to generate cyber attack contents for security exercises, labelling the attack scenario steps with ATT\&CK tactics.   

\subsubsection{⁠APT \& Malware Detection}

MITRE ATT\&CK framework is widely utilized in advanced persistent threats (APTs) detection, including attack modeling \cite{jabar2022mobile, soliman2023rank} and experimental evaluation \cite{park2023performance}. For instance, \cite{jabar2022mobile} models APT attack fingerprints using ATT\&CK TTP attack trees, while \cite{soliman2023rank} uses ATT\&CK techniques to map correlated alerts and construct an APT attack graph. The framework's attack group profiles and techniques are employed as data sources for evaluating APT detection models in \cite{park2023performance}. \cite{park2023performance} also uses the coverage percentage against ATT\&CK as a metric for APT detection performance. Beyond detection, ATT\&CK is also applied in APT attribution \cite{gonzalez2023technical, shin2021art, sachidananda2023apter}. For example, \cite{shin2021art} presents a threat actor classification method by comparing the TTPs of various APT groups, using ATT\&CK matrices extracted from cyber threat reports.

ATT\&CK is also extensively used in general malware detection. Several studies incorporate ATT\&CK with ransomware research \cite{ahn2023exploring, song2023similarity, singh2024s}. In \cite{ahn2023exploring}, APT scenarios based on ATT\&CK are used for evaluation. ATT\&CK is also correlated with ransomware data semantic features for detection \cite{singh2024s}. Malware attribution approaches, similar to APT attribution, profile malware to ATT\&CK tactics and techniques using NLP methods \cite{domschot2024improving}. Additional uses of ATT\&CK include playbook generation for malware deception \cite{sajid2023symbsoda}, and malware evolution analysis \cite{chierzi2021evolution}.

\subsubsection{Vulnerability Management}

Researchers analyze the vulnerabilities for effective threat prevention and security patching. With the incorporation of MITRE ATT\&CK, vulnerability management is often combined with risk assessment \cite{yoon2023vulnerability, devaux2021automation, muthia2023risk} and cross-data source analysis \cite{hacks2022multi, touloumis2021vulnerabilities, nazzal2022vulnerability}. Vulnerabilities are mapped to ATT\&CK TTPs in \cite{muthia2023risk} and \cite{touloumis2021vulnerabilities, nazzal2022vulnerability} for threat-integrated risk score computation and in-depth vulnerability analysis. In \cite{devaux2021automation}, ATT\&CK and cyber kill chain are used to construct operational risk scenarios linked to vulnerabilities for vulnerability-based risk management. Other vulnerability researches also leverage the linkage between ATT\&CK and vulnerability information, such as estimating attack time-to-compromise based on vulnerabilities \cite{rencelj2023estimating} and evaluating security robustness through vulnerability testing \cite{sikandar2022adversarial}.

\subsection{Industry-Specific Applications in Literature}
\label{sec:ApplicationDomain}

In addition to analyzing research focus areas, we examined the application of the MITRE ATT\&CK framework across various industries. Drawing from the 16 critical infrastructure sectors defined by CISA \cite{cisa2024critical}, we identified 12 key industry pillars encompassing critical assets, systems, and networks vital for maintaining societal and economic stability. To ensure comprehensive coverage, we include \textit{General} category for papers that lack a direct industry focus. Table~\ref{tab:industry} summarizes these industries, their descriptions, and associated research papers. This section highlights ATT\&CK's use in industries with at least five related papers, emphasizing unique applications rather than reiterating models and approaches discussed earlier. The \textit{General} category, accounting for 43.1\% of the papers (180 out of 417), reflects ATT\&CK’s broad applicability across domains but also indicates the need for more industry-specific applications. Sectors like Energy (9 papers, 2.2\%), Transportation Systems (10 papers, 2.4\%), Media (5 papers, 1.2\%), and Healthcare (5 papers, 1.2\%) are underrepresented.

\begin{table}[h]
\footnotesize
\caption{Industries of MITRE ATT\&CK Application.}
\label{tab:industry}
\resizebox{1\textwidth}{!}{%
\begin{tabularx}{\textwidth}{|X|l|l|}
\hline
\textbf{Industry Sector} & Frequency & \textbf{Examples} \\ \hline
\textbf{General}: Covers research not focused on any specific industry sector. & 174  & \cite{legoy2020automated, elitzur2019attack, straub2020modeling, straub2020modeling} \\ \hline

\textbf{Information Technology}: Hardware, software, and other assets in IT systems and services. & 124 & \cite{xiong2022cyber, blaise2022stay, dravsar2020session, mayukha2022reconnaissance} \\ \hline

\textbf{Operation Technology \& Manufacturing}: Hardware, software, and other assets that interact with the physical environment and manufacturing facilities. & 60 & \cite{song2024genics, mustafa2023cpgrid, yockey2023cyber, rencelj2023estimating} \\ \hline

\textbf{Communications}: Terrestrial, satellite, and wireless transmission systems. & 29 & \cite{pell2021towards, santos2022threat, simonetto2024strengthening, gulatas2023malware} \\ \hline

\textbf{Transportation Systems}: Transportation facilities and services, including aviation, highway, maritime, and other sectors. & 10 & \cite{amro2023evaluation, yousaf2024sinking, yekta2023vatt}  \\ \hline

\textbf{Energy}: Infrastructure of electricity, oil, and natural gas resources and assets to maintain energy supplies. & 9 & \cite{mccarty2023cybersecurity, cerotti2020evidence, ullah2019cyber, amro2023evaluation} \\ \hline

\textbf{Healthcare \& Public Health}: Businesses that manufacture and provision medical services, products, and insurance to public. & 5 & \cite{ampel2024improving, karagiannis2021demo, ahmed2022mitre, karagiannis2021demo} \\ \hline

\textbf{Media}: Services and tools that produce, store and deliver media contents. & 4 & \cite{yameogo2024improving, marinho2023automated, purba2023extracting} \\ \hline

\textbf{Government Services \& Facilities}: Services, Physical and cyber assets owned or leased by federal, state, local, and tribal governments. & 4 & \cite{mondal2023security, rajesh2022analysis, lilly2019applying, lanni2024boosting} \\ \hline

\textbf{Education}: Organizations and businesses that provide products and services of knowledge and skill learning. & 4 & \cite{bleiman2023exploring, canada2023recommendations, luh2022penquest, omiya2019secu} \\ \hline

\textbf{Defense Industrial Base}: Companies that develop, produce, deliver, and maintain military weapons systems and products. & 3 & \cite{batalla2024threat, Mohamed2022air, lee2022icstasy} \\ \hline

\textbf{Water \& Wastewater Systems}: Systems that provide clean drinking water supply and wastewater treatment. & 2 & \cite{alfageh2023water, toker2021mitre} \\ \hline

\textbf{Financial Services}: Depository institutions, investment providers, insurance companies, and other financing organizations. & 1 & \cite{alevizos2023cyber} \\ \hline

\end{tabularx}}
\end{table}

\subsubsection{Information Technology, Operation Technology \& Manufacturing, Communications}

The categorization of ATT\&CK-related papers into the IT, OT \& manufacturing, and communications sectors is based on the types of cyber systems and assets investigated. Table~\ref{tab:attck_tech_application} summarizes the applications of the MITRE ATT\&CK framework across four domains, highlighting the cyber systems in each sector. While there is significant overlap in how ATT\&CK is utilized across different cybersecurity systems, the specific ATT\&CK matrix employed often varies. The framework’s diverse applications are prominently seen in enterprise IT networks, OT industrial control systems, and 5G/IoT communication environments.  

\begin{table}[h]
\footnotesize
\caption{ATT\&CK application summary across IT, OT and communications cybersecurity systems.}
\label{tab:attck_tech_application}
\begin{tabular}{|l|l|l|l|}
\hline
\textbf{Application} & \textbf{IT} & \textbf{OT} & \textbf{Communications} \\ \hline
\begin{tabular}[c]{@{}l@{}}Threat model \\ component\end{tabular} & \begin{tabular}[c]{@{}l@{}}Enterprise IT system \cite{xiong2022cyber}\\ Self-sovereign identity system \cite{naik2022evaluation}\end{tabular} & \begin{tabular}[c]{@{}l@{}}ICS \cite{song2024genics, zhang2022threat}\\ Powerstation control systems \cite{horalek2023security}\end{tabular} & \begin{tabular}[c]{@{}l@{}}5G networks \cite{pell2021towards,  santos2022threat, vanderveen2022threat} \\ Mobile networks \cite{rao2023threat, van2023mitre}\end{tabular} \\ \hline
\begin{tabular}[c]{@{}l@{}}Attack scenario \\ generation \end{tabular} & \begin{tabular}[c]{@{}l@{}}IT network attack simulation \cite{dravsar2020session} \\ SOC assessment \cite{rosso2022saibersoc} \\ Penetration testing \cite{mayukha2022reconnaissance}\end{tabular} & \begin{tabular}[c]{@{}l@{}}ICS attack data generation \cite{choi2021probabilistic, choi2020expansion}\\ CPS \cite{serru2022modeling}\\ Power grid \cite{mustafa2023cpgrid}\end{tabular} & 5G, IoT, IIoT \cite{simonetto2024strengthening} \\ \hline
\begin{tabular}[c]{@{}l@{}}Security \\ assessment\end{tabular} & \begin{tabular}[c]{@{}l@{}}Enterprise networks \cite{manocha2021security, he2021model}\\ Cloud and containers \cite{blaise2022stay}\end{tabular} & \begin{tabular}[c]{@{}l@{}}ICS evaluation \cite{wang2023rrdd}\\ Autonomous control systems \cite{yockey2023cyber}\end{tabular} & IIoT \cite{falco2018master} \\ \hline
\begin{tabular}[c]{@{}l@{}}Threat \\ categorization \end{tabular} & \begin{tabular}[c]{@{}l@{}}Malware detection \cite{ahn2022malicious}\\ NIDS \cite{wang2022ai}\end{tabular} & ICS \cite{rencelj2023estimating, nursidiq2022threat, ekisa2024leveraging} & IoT malware \cite{gulatas2023malware, chierzi2021evolution} \\ \hline
\end{tabular}
\end{table}

\subsubsection{Transportation Systems}

In the transportation industry, the MITRE ATT\&CK framework is primarily applied to maritime and railroad security. 

In maritime cyber risk assessment, studies such as \cite{amro2023evaluation, amro2023cyber} focus on autonomous passenger ships, integrating ATT\&CK to enhance existing risk assessment methodologies. Similarly, \cite{oruc2022assessing} addresses the cyber risks of marine components within integrated navigation systems, modifying traditional risk assessment frameworks to include ATT\&CK TTPs and mitigation strategies. ATT\&CK framework is also employed in maritime threat modeling. For example, \cite{yousaf2024sinking} uses a case study involving an attack on a ship’s ballast water management system to model threats using the ATT\&CK matrix. This study further proposes defense strategies derived from ATT\&CK mitigations and the MITRE D3FEND framework, showcasing the practical utility of MITRE tools. Similarly, \cite{jo2022cyberattack} presents an attack model for shipping equipment that leverages ATT\&CK TTPs to characterize attack stages based on vulnerability and threat research.

The application of MITRE ATT\&CK framework in railroad systems is observed in \cite{yekta2023vatt, kim2023securing}. \cite{yekta2023vatt} introduces the \textsc{VATT\&EK} framework, which categorizes common cyber threats in intelligent transport systems using ATT\&CK TTPs, formalizing vehicle-related attacks and supporting threat prioritization. Meanwhile, \cite{kim2023securing} applies the ATT\&CK matrix to generate attack sequence diagrams for evaluating the resilience of a proposed blockchain-based non-stop customs clearance (NSCC) system, designed to reduce resource requirements and delays in railroad operations.

\subsubsection{Energy}

ATT\&CK framework, specifically the ICS matrix, is applied in the energy sector for cyber resilience assessments. Studies like \cite{mccarty2023cybersecurity, gourisetti2022assessing} leverage ATT\&CK to evaluate the cyber resilience of energy systems. \cite{mccarty2023cybersecurity} performs cost-benefit analysis and risk assessments of security controls on cyber-physical wind energy sites, using the ATT\&CK ICS matrix to generate attack scenarios for both remote and local energy system network access. Similarly, \cite{gourisetti2022assessing} evaluates Distributed Ledger Technology (DLT) in the energy sector, proposing a cyber resilience model where TTPs are mapped to various DLT stack layers for risk analysis and mitigation recommendations.

ATT\&CK framework also supports cyber threat modeling for energy systems. For example, \cite{cerotti2020evidence} develops an attack knowledge base by mapping publicly reported attack data to ATT\&CK ICS TTPs in distributed energy control systems. The study models these scenarios as attack graphs with attacker state nodes and attack technique edges. Dynamic Bayesian Networks derived from these graphs capture causal and temporal dependencies of threats, enabling detailed risk assessments. Similarly, \cite{ullah2019cyber} identifies threat indicators and network topology in energy delivery systems to construct TTP-based attack graphs. These models assess the likelihood of asset compromise along attack paths using defined risk metrics.

In the oil and gas sector, field flooding attacks targeting the memory structure of Modbus operations in programmable logic controllers (PLCs) have become a growing concern. \cite{amro2023evaluation} uses the ATT\&CK framework to assess flooding attacks on an industrial testbed. This study identifies the associated ATT\&CK TTPs, evaluates the attack methodology and impacts, and proposes a tailored detection algorithm to mitigate such threats.

\subsubsection{Healthcare \& Public Health}

Research in the healthcare sector primarily focuses on securing healthcare infrastructures and medical devices. For instance, ATT\&CK is employed for threat and attack route identification in the security analysis of radiological medical devices, with mitigation strategies proposed using the ATT\&CK knowledge base \cite{zisad2024towards}. Similarly, \cite{ampel2024improving} introduces the \textsc{ATT\&CK-Link} framework, which utilizes a transformer-based model and multimodal classification to link hacker threats in hospital systems to ATT\&CK. This approach generates actionable CTI for comprehensive security analysis in healthcare infrastructure.

Threat emulation provides a means to assess adversarial behavior and evaluate the effectiveness of security controls. \cite{karagiannis2021demo} proposes \textsc{A-DEMO}, a structured attack emulation methodology that models adversarial techniques and mitigation plans using ATT\&CK. A case study on a replicated healthcare infrastructure demonstrates the methodology’s effectiveness through the emulation of a rootkit attack. Additionally, \cite{ahmed2022mitre} integrates ATT\&CK TTPs and attack graphs to evaluate attack success likelihood. High-likelihood attack paths, represented as TTP sequences, are extracted for deeper analysis. Simulations of APT attacks by Lazarus and menuPass groups targeting healthcare organizations highlight the approach's validity, with risk scoring conducted under varying security control coverage levels.

\subsubsection{Media}

Media platforms, particularly social media, serve as a rich source for CTI research due to the diversity and volume of shared information. Studies such as \cite{marinho2023automated, purba2023extracting} leverage social media posts as input for threat detection and profiling. \cite{marinho2023automated} employs Twitter messages to identify threats and their associated names using a binary classifier trained on the ATT\&CK corpus. A multi-class ML classifier, trained on ATT\&CK procedures grouped by tactics, profiles the detected threats. Similarly, \cite{purba2023extracting} proposes an ATT\&CK-based cyberattack pattern extraction method from tweets. Semantic filtering compares phrases extracted by Semantic Role Labeling (SRL) from tweets with those extracted from the ATT\&CK corpus. Extracted terms, tags, and IoCs are then mapped to ATT\&CK techniques and tactics, generating technical threat profiles.

Although media is useful for CTI extraction and analysis, attackers can also leverage the vulnerabilities and indistinguishable misinformation in media applications to perform adversarial actions. The defined attack life cycle based on ATT\&CK framework has 4 stages, including reconnaissance, initial compromise, command and control, and exfiltration. Assuming success of the first 2 stages, the feasibility of social media as C2 channel is researched and a case study is conducted, proving Twitter can communicate with victims and evade existing detections as a C2 channel. \cite{yameogo2024improving} proposed a two-stage methodology against informational attacks, which include various fake news on social media platforms. After entities are extracted from fake news by NLP techniques and uniformly formatted, in-depth analysis is performed, including clustering and factorial analysis, to discover the disinformation campaigns. identified compaigns can be mapped to the Disinformation Analysis and Response Measures (DISARM) framework of MITRE ATT\&CK framework TTP styling for subsequent analysis. 

\subsubsection{General}

Research in the general cybersecurity domain applies broad methodologies without targeting specific industries. The MITRE ATT\&CK framework is widely applied across various research areas, serving as a foundational model for analyzing cyber threats and enabling systematic categorization. In CTI, ATT\&CK TTPs are frequently used to structure and profile threat patterns from data sources such as vulnerability reports, attack narratives, and IDS events \cite{legoy2020automated, ayoade2018automated, grigorescu2022cve2att}. For threat hunting, ATT\&CK functions as an evaluation framework \cite{nisioti2021game} or knowledge base \cite{elitzur2019attack} for generating attack hypotheses and conducting cyber forensics. Studies have mapped network incidents, malware, and APT data to ATT\&CK TTPs for comprehensive attack characterization \cite{aghamohammadpour2023architecting, gonzalez2023technical}. In threat modeling and attack simulation, ATT\&CK is often integrated into frameworks to characterize threats and construct attack scenarios. For instance, \cite{straub2020modeling, franklin2017toward} utilize the ATT\&CK matrix in threat models, while \cite{takahashi2020aptgen, hacks2021integrating} define ATT\&CK-based attack sequences for simulation studies. Additionally, ATT\&CK has been applied for mapping attack actions to security control assessments \cite{rahman2022investigation} and supporting risk analysis \cite{belfadel2022towards}.

\section{Data Sources and Collection Tools} \label{sec:DataSources}

The integration of tools and dataset with the MITRE ATT\&CK framework is central to advancing cybersecurity research and enhancing threat detection capabilities. This section categorizes and analyzes the tools, data sources, and techniques utilized in MITRE ATT\&CK-based research.

\subsection{Data Collection Tools}

Tools for collecting cyber threat data and aligning it with MITRE ATT\&CK framework include:

\textbf{Threat Intelligence and Adversary Simulation Tools} are vital for understanding and countering malicious activities, offering capabilities to aggregate intelligence, simulate attacks, and analyze attacker behaviors. Platforms like Malware Information Sharing Platform (MISP)  \cite{mundt2024enhancing, ammi2023cyber, parmar2019use} help aggregate and share threat intelligence and analyze potential threat indicators. Tools such as Atomic Red Team \cite{okada2024predicting, orbinato2024laccolith, bierwirth2024design} and Caldera \cite{chetwyn2024modelling, lee2023codex, noel2023graph, gjerstad2022lademu} simulate adversary TTPs that are further mapped to MITRE ATT\&CK. Honeypots and deception frameworks are designed to attract attackers and analyze their behaviors. Particularly, Cowrie Honeypot \cite{lin2024hybrid, mohammadzad2024cyber} logs detailed interactions with attackers, while Honeytrap \cite{subhan2023analyzing} captures and studies malicious activities. GridPot \cite{izzuddin2022mapping} emulates ICS, while Conpot \cite{arafune2022design} provides a lightweight and configurable honeypot for ICS environments. Simulation tools and testbeds create controlled environments for evaluating security measures against realistic threats. GridEx \cite{mustafa2023cpgrid} provides large-scale simulation of cyber threats, for instance. The EPIC \cite{mumrez2023comparative} Testbed supports experimentation with adversary tactics, and Factory IO \cite{mekala2023attacks} offers an industrial automation environment for testing.

\textbf{Vulnerability Assessment, Penetration Testing, and Monitoring Tools} provide critical insights into system weaknesses, which enables proactive threat detection and response. To be more specific, Nmap, Nessus, and OpenVAS are used for network scanning and vulnerability discovery. Wireshark \cite{maesschalck2024these, houmb2023intelligent, alsabbagh2024investigating, maesschalck2024these} and SNORT \cite{lin2024hybrid, afenu2024industrial, daniel2023labeling} monitor network traffic and detect suspicious patterns. Burp Suite \cite{mayukha2022reconnaissance} and Metasploit \cite{rencelj2023estimating, yoon2023vulnerability, mekala2023attacks, ajmal2023toward} provide penetration testing capabilities, simulating real-world exploitation scenarios. Security Information and Event Management (SIEM) \cite{aragones2023threat, cakmakcci2023central, bryant2020improving} and other monitoring tools such as Splunk \cite{virkud2024does, bierwirth2024design, noel2023graph, abhan2022threat, parmar2019use}, Logstash \cite{marinho2023automated, marinho2023framework, kulkarni2023proactive}, IBM QRadar \cite{aragones2023threat}, Kibana \cite{youn2022research, hermawan2021development}, System Monitor (Sysmon) \cite{rodriguez2024process, okada2024predicting, chetwyn2024modelling}, Sigma Rules \cite{rodriguez2024process, chetwyn2024modelling, rodriguez2023discovering}, Security Onion \cite{bagui2023introducing},and Wazuh \cite{ammi2023cyber, amro2023cyber, touloumis2022tool} aggregate, parse, and analyze logs and events to detect and respond to security threats. VirusTotal \cite{leite2022actionable, shin2021art, gulatas2023malware} specifically facilitatez the identification and characterization of malware samples. For example, \cite{maesschalck2024these} uses Wireshark for passive discovery of network devices, comparing traffic behavior between obfuscated and non-obfuscated PLCs to assess network-level obfuscation. Similarly, \cite{houmb2023intelligent} employs Wireshark to analyze network traffic in Hardware-in-the-Loop (HIL) simulations. \cite{ammi2023cyber} proposes a four-layered architecture that integrates multiple tools like MISP, Wazuh, IDS/IPS, and Elastic Stack, a Security Information and Event Management (SIEM) platform. More specifically, The MISP platform serves as the central hub for collecting, analyzing, and disseminating threat intelligence data. Then Wazuh alongside Sysmon to capture detailed logs from the Windows operating system. 

\textbf{Endpoint Detection and Response (EDR) Tools} leverage advanced techniques to detect and respond to sophisticated attack pattern. Tools like CrowdStrike \cite{ajmal2023toward}, VMware Carbon Black \cite{patil2023audit, virkud2024does} and Microsoft Defender \cite{orbinato2024laccolith} monitor endpoints for malicious behaviors aligned with MITRE ATT\&CK and enhance detection of sophisticated attack patterns. The Laccolith system \cite{orbinato2024laccolith}, for instance, utilizes Microsoft Defender to simulate adversary techniques, evaluating their ability to evade detection while refining detection methods. \cite{kulkarni2023proactive} proposes a proactive framework for advanced threat hunting, emphasizing the role of EDR solutions in identifying and mitigating undetected threats.

\subsection{Data Sources}

Common datasets and logs used for MITRE ATT\&CK-based applications include:

\textbf{Logs and Network Traffic Data} provide insights into system activities and network behaviors. Examples include Syslog \cite{mustafa2024threat, aragones2023threat}, Zeek logs \cite{lee2023codex, bagui2023introducing, alkhpor2023collaborative}, and traffic capture tools like SPAN \cite{toure2024framework}, NetFlow \cite{toure2024framework, choi2020expansion}, and PCAPs \cite{sharma2023radar, sen2023approach, bagui2023introducing, aragones2023threat}, which help detect anomalies and map attack vectors. SIEM logs and firewall or VPN logs aggregate this data to enhance monitoring and response capabilities. For example, \cite{luo2024detecting} relies on system audit logs as its primary data source to provide a detailed record of system events, capturing interactions between various entities such as processes, files, and sockets. \cite{okada2024predicting} employs Windows Sysmon logs to extract ATT\&CK techniques. \cite{choi2020expansion} collects a comprehensive dataset for security research in ICS by expanding a Hardware-In-The-Loop (HIL) based augmented ICS (HAI) testbed, including network traffic, NetFlow data, system event and security appliance logs, control system (e.g., PLC) logs, and simulated attack data. 
    
\textbf{Public Datasets} are commonly used in MITRE ATT\&CK research. Vulnerability and incident databases like NVD/CVE \cite{grigorescu2022cve2att, ampel2023mapping, nazzal2022vulnerability}, ExploitDB \cite{ampel2023mapping, hemberg2024enhancements}, and Metasploit DB \cite{rencelj2023estimating, yoon2023vulnerability, mekala2023attacks} catalog known vulnerabilities and exploits. Meanwhile, APT reports (e.g., FireEye APT1 report \cite{ajmal2023toward, shin2021art}), malware reports (e.g., MalwareBazaar \cite{gonzalez2023technical, gulatas2023malware}) and government publications (e.g., ENISA \cite{rostami2020machine, lakhdhar2021machine}, MITRE Cyber Analytics Repository (CAR) \cite{cerotti2020evidence, hemberg2024enhancements}, and CERT Advisories \cite{shin2021art, cerotti2020evidence, cerotti2019analysis}) provide invaluable insights into past cyber incidents and adversary behaviors. For example, \cite{ampel2023mapping} uses CVEs to identify vulnerabilities associated with the collected 29,259 exploit source code samples written in Python from Pastebin and 1,959 exploits from ExploitDB. \cite{ajmal2023toward} employs FireEye’s CTI reports to extract attack details and behavioral insights related to the APT41 threat group, then maps high-level attack tactics and techniques to low-level system event logs for enhanced detection and analysis of APT activities. \cite{cerotti2020evidence} uses security analytics from MITRE CAR to model aspects related to attack detection, and derives scores and parameters for their models from ICS-CERT advisories. \cite{rencelj2023estimating} utilizes a ICS-specific vulnerability dataset, Metasploit database and other external reports for estimating the Time-To-Compromise (TTC) for different ICS attack techniques defined in the MITRE ATT\&CK framework.
    
\textbf{Open-Source Project Datasets} play a critical role in understanding cyber threats. Key datasets include the CIC-IDS collections \cite{arreche2024xai, arreche2024two, kim2024relative, borisenko2022cse}, which offer labeled network traffic data simulating diverse attack scenarios like DoS, brute force, and botnet attacks. These datasets are widely used for benchmarking intrusion detection models. The NSL-KDD dataset \cite{arreche2024two, preethi2023leveraging, toure2024framework} provides improved versions of the original KDD Cup 1999 dataset by eliminating duplicate records, making it suitable for evaluating anomaly detection algorithms. The DARPA datasets span multiple research initiatives, including DARPA TC3 Theia \cite{luo2024detecting} that comprise Linux system audit logs collected during a red-team vs. blue-team adversarial engagement in April 2018, DARPA Transparent Computing \cite{dong2023distdet, li2022attackg, patil2023audit} that focuses on collecting and analyzing provenance data, system logs, and network traffic to study APTs and system-level compromises, and also DARPA OpTC \cite{zipperle2024pargmf, patil2023audit} that offers data from cyber-physical systems, with a focus on OT environments. Specialized datasets, such as RoEduNet-SIMARGL2021 \cite{arreche2024two}, focus on IoT threats by capturing malicious activities targeting smart devices and networks. This dataset is particularly valuable for developing IoT-specific intrusion detection systems. VERIS Community Database (VCDB) \cite{dev2023models} aggregates information from publicly disclosed cybersecurity incidents, such as breaches reported in media or security reports. There are also some other open-source intelligence (OSINT) \cite{ammi2023cyber, aragones2023threat, youn2022research, gulatas2023malware} that enhances adversary profiling. For instance, \cite{arreche2024two} evaluates the proposed two-level ensemble learning framework on three network intrusion datasets (RoEduNet-SIMARGL2021, CICIDS-2017, NSL-KDD) that include network traffic representing different attack types, and utilizes the MITRE ATT\&CK framework as a reference point to categorize and understand different network intrusion types. \cite{youn2022research} integrates OSINT with BGP archive data to improve the identification and tracking of cyberattack origins, particularly for North Korean threat actors. \cite{gulatas2023malware} collects OSINT data from malware databases, academic papers, and antivirus reports to analyze IoT malware behaviors.

\section{Integrated Framework and Analytical Tools} \label{sec:AnalyseTools}

This section explores how cybersecurity frameworks and analytical tools are utilized in the reviewed papers to enhance threat detection, incident response, and system resilience. 

\subsection{Integration with Other Frameworks} \label{sec:Framework}

Key frameworks and their applications include:

\textbf{Cybersecurity Frameworks and Standards} provide standardized guidelines and practices to ensure a comprehensive and structured approach to managing cyber risks. Key examples include the NIST frameworks \cite{chimmanee2024digital} (e.g., NIST SP 800-30 \cite{dev2023models, ahmed2022mitre}, NIST SP 800-53 \cite{kern2024logging, derbyshire2021talking, chorfa2023threat, bodeau2022using}, NIST SP 800-172 \cite{bodeau2022using}, NIST SP 800-154 \cite{alevizos2023cyber}, NIST SP 800-207 \cite{tsai2024strategy} and NIST SP 800-61 \cite{wilson2024multi}), which emphasizes the identification, protection, detection, response, and recovery lifecycle, as well as the ISO/IEC frameworks \cite{horalek2023security} (e.g., ISO 27001 \cite{belfadel2022towards, mundt2023threat}, ISO 27002 \cite{derbyshire2021talking}, ISO 12100 \cite{brenner2023better}, IEC 62443 \cite{brenner2023better, horalek2023security}, IEC 61508 \cite{brenner2023better}), which focus on information security management systems. The CIS Critical Security Controls (CSC) \cite{kern2024logging, derbyshire2021talking} offers prioritized actions to mitigate threats, while the Purdue Enterprise Architecture \cite{houmb2023intelligent, horalek2023security} and RAMI 4.0 \cite{masi2023securing} provide models specifically tailored to industrial control systems and OT. Facility Cybersecurity Framework (FCF) \cite{kwon2020cyber} integrates structured controls with governance requirements. Additional standards such as ISA 18.2, API 1167 and EEMUA 191 provide specific guidance for process safety and alarm management \cite{alabdulhadi2023alarm}.  In \cite{kern2024logging}, for example, CIS Controls are integrated into a logging maturity model to aid organizations in selecting IDS solutions while balancing constraints like cost and complexity. Additionally, \cite{derbyshire2021talking} explores the use of SANS CIS Controls alongside MITRE ATT\&CK techniques but ultimately highlights NIST 800-53 for its granular control layers, demonstrating the flexibility of CIS Controls in adapting to specific needs. \cite{tsai2024strategy} integrates elements from NIST SP 800-207 and MITRE ATT\&CK framework into a Zero Trust Architecture (ZTA). 

\textbf{Threat and Attack Modeling Frameworks} are essential for understanding adversary behaviors. Frameworks such as STRIDE \cite{davis2024cyber, mundt2024enhancing, alevizos2023cyber} and D3FEND \cite{yousaf2024sinking, hemberg2024enhancements, kim2023design} PASTA \cite{dev2023models} focus on identifying threats in software and systems, while Diamond Model \cite{nursidiq2023cyber, mckee2023activity, kulkarni2023proactive, hermawan2021development} and Cyber Kill Chain \cite{omiya2019secu, lilly2019applying, straub2020modeling, falco2018master, perez2022evaluation} emphasize attack progression and the relationship between attacker objectives and defender responses. Some works include the Pyramid of Pain \cite{hermawan2021development}, Markov Models \cite{choi2021probabilistic, atefi2023principled}, and attack trees \cite{straub2020modeling, gupta2022comparative}, which provide structured approaches for visualizing and analyzing attack paths. Meanwhile, simulation models such as securiCAD \cite{hacks2022multi, hacks2021integrating, gylling2021mapping}, Prolog \cite{liu2020forensic, venkatesan2019vulnervan}, SyncAD \cite{mustafa2023cpgrid}, GridEx \cite{mccarty2023cybersecurity, mustafa2023cpgrid}, Meta Attack Language (MAL) \cite{hacks2021integrating, gylling2021mapping, masi2023securing} are used to simulate adversarial behaviors and system responses. For instance, \cite{thein2020paragraph} extracts event information from security reports and applies Cyber Kill Chain to categorize attack phases. \cite{davis2024cyber} utilizes STRIDE to classify threats to Water and Wastewater Systems (WWS) and evaluates their risk using DREAD. In \cite{kim2023design}, D3FEND is integrated into a defense strategy against IoT botnet attacks within smart city environments. \cite{sanchez2023analysis} analyzes and compares multiple frameworks, specifically the Diamond Model, Cyber Kill Chain, and ATT\&CK, for AI data processing and categorization of attack tactics.

\textbf{Intelligence Sharing Frameworks} such as Structured Threat Information Expression (STIX) \cite{marchiori2023stixnet, mckee2023activity, golushko2022adversary}, Trusted Automated Exchange of Indicator Information (TAXII) \cite{marchiori2023stixnet}, and Cyber Observable eXpression (CyBox) (briefly mentioned in \cite{Kurniawan2021AnAF}), enable organizations to standardize and exchange threat information. For instance, \cite{marchiori2023stixnet} introduces STIXnet, a tool that utilizes TAXII to retrieve data from the MITRE ATT\&CK framework and then employs a ML model to recognize tactics and techniques that are not explicitly mentioned in CTI reports. Similarly, \cite{golushko2022adversary} recommends utilizing the STIX 2.x standard when building adversary profiles, particularly on classifying an adversary's skill level.

\textbf{Vulnerability Assessment Standards} enhance the applicability of the ATT\&CK framework by mapping vulnerabilities to adversary tactics and techniques. Key examples include the Common Vulnerability Scoring System (CVSS) \cite{wang2023rrdd, yoon2023vulnerability, chorfa2023threat, muthia2023risk, oruc2022assessing}, which evaluates the severity of vulnerabilities based on exploitability and impact, and the Common Weakness Scoring System (CWSS), which assesses weaknesses in software architecture. Complementary systems like Common Attack Pattern Enumeration and Classification (CAPEC) \cite{irshad2023cyber, falco2018master, rostami2020machine, omiya2019secu}, Common Weakness Enumeration (CWE) \cite{villanueva2023analyzing, lakhdhar2021machine, touloumis2021vulnerabilities, purba2020word} , and the Security Qualification Matrix (SQM) \cite{manocha2021security} provide detailed categorizations of attack patterns and weaknesses. For example, studies like \cite{ampel2023mapping, hemberg2024enhancements} correlate CVE reports with ATT\&CK techniques through CWE and CAPEC. Similarly, \cite{omiya2019secu} employs Cyber Kill Chain to structure attack stages and categorize attack methods derived from MITRE ATT\&CK and CAPEC. \cite{villanueva2023analyzing} demonstrates the use of NLP techniques, including TF-IDF, BERT, and SBERT, to map CWEs to corresponding ICS ATT\&CK techniques based on their textual descriptions.

\subsection{Analytical Tools and Techniques} \label{sec:Analytical}

Tools and techniques for processing and analyzing data include:

\textbf{Machine Learning (ML) and Natural Language Processing (NLP) Tools} are integrated with the MITRE ATT\&CK framework to analyze large datasets to identify patterns and map them to adversary tactics, as discussed in \ref{sec:CTI}. ML models such as XGBoost \cite{mohammed2023detection, alkhpor2023collaborative, guha2023evaluation}, Random Forest (RF) \cite{alkhpor2023collaborative, sharma2023ttp, irshad2023cyber, takey2022real}, Logistic Regression \cite{takey2022real}, Support Vector Machine (SVM) \cite{ampel2023mapping, sachidananda2023apter, brenner2023better, grigorescu2022cve2att}, Decision Tree (DT) \cite{sharma2023ttp, irshad2023cyber}, DNN \cite{yi2024anomaly, arreche2024two}, CNN \cite{nguyen2024noise, sun2023sectkg, grigorescu2022cve2att}, and Long Short Term Memory (LSTM) \cite{xiao2024research, domschot2024improving, patil2023audit, sun2023sectkg}, then transformer-based models such as Bidirectional Encoder Representations from Transformers (BERT) \cite{singh2024s, yameogo2024improving, lin2024hybrid, hemberg2024enhancements, huang2024mitretrieval} and its variances (e.g., RoBERTa \cite{rani2023ttphunter, alam2023looking}, DistilBERT \cite{li2024automated, ge2024metacluster}, and SecBERT \cite{alves2022leveraging, nguyen2024noise}) and recently emerged language models like Generative Pre-trained Transformer (GPT) \cite{branescu2024automated, song2023generating}, are increasingly used in cybersecurity for threat detection and response. For example, studies such as \cite{lakhdhar2021machine, lin2024hybrid, nguyen2024noise} propose multi-label classification models to associate vulnerabilities or other textual reports with adversarial tactics. In \cite{alves2022leveraging}, eleven different BERT models, including pre-trained models like BERT Base, BERT Large, RoBERTa, DistilBERT, and models pre-trained on cybersecurity text like SecBERT and SecRoBERTa, are fine-tuned and evaluated on their abilities to identify and categorize TTPs from procedure examples of MITRE ATT\&CK. BERT models with larger number of parameters generally achieve better accuracies. \cite{branescu2024automated} also demonstrates that transformer-based models can effectively automate the mapping of CVEs to MITRE ATT\&CK tactics. In this study, SecRoBERTa achieves the best overall performance, while GPT-4, despite its zero-shot learning capabilities, exhibits significantly lower performance. Nevertheless, GPT models demonstrate their usefulness in labeling SNORT rules with MITRE ATT\&CK techniques in \cite{daniel2023labeling}. Meanwhile, NLP tools like GloVe \cite{domschot2024improving, hemberg2024enhancements}, TF-IDF \cite{tang2023attack, liao2023intelligent}, Word2Vec \cite{sajid2023symbsoda, sun2023sectkg, andrew2022mapping}, and SpaCy \cite{zacharis2023aicef} assist in extracting intelligence from unstructured data.

\textbf{Graph-Based and Knowledge Representation Tools} often integrate with the MITRE ATT\&CK framework to model attack paths and identify critical nodes in a system. Neo4j \cite{edie2023extending, kaiser2022cyber}, GraphSAGE \cite{bagui2024graphical} and CyGraph \cite{noel2023graph} support the creation of attack graphs, mapping adversary behaviors and system vulnerabilities to MITRE ATT\&CK techniques. Ontology tools like Web Ontology Language (OWL) \cite{huang2022building, Kurniawan2021AnAF}, Resource Description Framework (RDF) \cite{akbar2023design} and SPARQL Protocol and RDF Query Language (SPARQL) \cite{chetwyn2024modelling, heverin2023reconnaissance, akbar2023design} further enhance knowledge sharing and querying capabilities. For example,  Neo4j is used in \cite{edie2023extending} to visually represent the relationships between APT groups and their TTPs in a KG format. \cite{akbar2023design} explains how to build and query a unified cybersecurity ontology that integrates various data sources like MITRE ATT\&CK, D3FEND, ENGAGE, CWE, and CVE, using RDF and SPARQL.

\textbf{Visualization Tools} play a critical role in presenting complex attack patterns and dependencies. Tools such as Matplotlib \cite{arreche2024xai}, Gephi \cite{heverin2023reconnaissance}, and GraphStream \cite{bagui2023using} support graphical representation of attack paths and dependencies. Similarly, MITRE ATT\&CK Navigator \cite{van2023mitre, chorfa2023threat} facilitates the organization, mapping, and analysis of adversary TTPs. For example, \cite{chorfa2023threat} integrates the Navigator with the TRAM tool to map Software-Defined Virtual Network (SDVN) attack vectors to the ATT\&CK framework.

\section{Research Trends on Tactics, Techniques, and Procedures (TTPs)} \label{sec:TTPs}

This section analyzes the TTPs reported in the reviewed literature, highlighting their frequency, trends, and application across various domains. Table \ref{tab:merged_tactics} lists researches of tactical frequencies across Enterprise, Mobile, and ICS domains, stratified by attack lifecycle stages, and only listed the distributions from 2019 to 2024. Note that only explicitly studied tactics are included, excluding those merely mentioned or referenced. This table demonstrates a temporal shift, with most tactical representations experiencing substantive scholarly attention predominantly after 2020. This chronological pattern suggests an emergent academic and operational interest in structured threat intelligence taxonomies. The presented tactics, conceptualized as high-level strategic objectives in adversarial operations, offer perspectives on cybersecurity threat landscapes.

\begin{table}[ht]
\centering
\footnotesize
\caption{MITRE ATT\&CK Tactics Frequency by Year (2019–2024)}
\label{tab:merged_tactics}
\begin{tabular}{|l|c|c|c|c|c|c|c|c|c|l|}
\hline
\textbf{Tactic Name (ID)} & \textbf{2019} & \textbf{2020} & \textbf{2021} & \textbf{2022} & \textbf{2023} & \textbf{2024} & \textbf{Total} & \textbf{Examples} \\ \hline
Reconnaissance (TA0043)        & 0 & 0 & 2 & 9 & 23 & 16 & 50  &  \cite{landauer2022maintainable, bagui2022detecting, yi2024anomaly} \\ \hline
Resource Development (TA0042)  & 0 & 0 & 2 & 6 & 12 & 12 & 32  &  \cite{belfadel2022towards, liao2021generating, youn2022research}\\ \hline
Initial Access (TA0001)        & 2 & 4 & 15 & 22 & 39 & 27 & 109  &  \cite{varlioglu2024pulse, park2022performance, mustafa2024threat} \\ \hline
Execution (TA0002)             & 3 & 5 & 14 & 20 & 33 & 24 & 99  &  \cite{sharma2023ttp, lewis2024effects, perez2022evaluation} \\ \hline
Persistence (TA0003)           & 2 & 5 & 12 & 17 & 32 & 22 & 90  &  \cite{lee2022icstasy, takey2022real, park2023performance} \\ \hline
Privilege Escalation (TA0004)  & 2 & 4 & 10 & 18 & 29 & 21 & 81  &  \cite{shin2021art, chimmanee2024digital, amro2023cyber}\\ \hline
Defense Evasion (TA0005)       & 1 & 5 & 13 & 16 & 28 & 21 & 84  &  \cite{ajmal2021offensive, gylling2021mapping, amro2022click}\\ \hline
Credential Access (TA0006)     & 0 & 5 & 11 & 15 & 27 & 23 & 81  & \cite{kryukov2022mapping, varlioglu2024pulse, nisioti2021game} \\ \hline
Discovery (TA0007)             & 3 & 6 & 12 & 21 & 37 & 28 & 107  & \cite{sharma2023ttp, gulatas2023malware, lee2023game} \\ \hline
Lateral Movement (TA0008)      & 3 & 6 & 12 & 15 & 37 & 22 & 95  &  \cite{gourisetti2022assessing, song2023similarity, ccakmakcci2023apt}\\ \hline
Collection (TA0009)            & 1 & 5 & 9 & 12 & 27 & 20 & 74  &  \cite{chimmanee2024digital, batalla2024threat, lee2023game} \\ \hline
Command and Control (TA0011)   & 2 & 3 & 9 & 16 & 33 & 21 & 84   & \cite{amro2022click, yi2024anomaly, lee2023game} \\ \hline
Exfiltration (TA0010)          & 2 & 5 & 9 & 9 & 25 & 20 & 70  &  \cite{mundt2024enhancing, al2023analysis, landauer2022maintainable}\\ \hline
Impair Process Control (TA0106)  & 0 & 0 & 1 & 2 & 9 & 3 & 15   &  \cite{mumrez2023comparative, lee2023assessment, ekisa2024leveraging}\\ \hline
Inhibit Response Function (TA0107)  & 0 & 0 & 0 & 2 & 7  & 3  & 12   &  \cite{gourisetti2022assessing, liao2021generating, hotellier2024standard} \\ \hline
Impact (TA0040)                & 0 & 5 & 8 & 12 & 35 & 18 & 75  &  \cite{pell2021towards, lasky2023machine, hotellier2024standard}\\ \hline
\end{tabular}
\end{table}

Initial Access (TA0001), Execution (TA0002), and Discovery (TA0007) remain the most frequently reported tactics, collectively accounting for over 30\% of total mentions. This indicates a strong research emphasis on early and middle phases of adversarial campaigns, where attackers gain entry, execute payloads, and gather system information. Lateral Movement (TA0008) and Persistence (TA0003) are also prominent, reflecting the need to study how attackers navigate networks and maintain footholds post-infiltration. Command and Control (TA0011) demonstrates steady growth, suggesting a heightened focus on adversary communication channels. Interestingly, tactics like Resource Development (TA0042) and ICS-specific tactics, such as Impair Process Control (TA0106) and Inhibit Response Function (TA0107), remain underrepresented. Impair Process Control (TA0106) and Inhibit Response Function (TA0107) appear only after 2020, peaking in 2023, highlighting the emerging attention to vulnerabilities in ICS environments. Meanwhile, Privilege Escalation (TA0004) and Impact (TA0040) show consistent increases, reflecting the ongoing research into advanced adversarial strategies and attack consequences.

\subsection{Early Attack-Stage Tactics} 

This section discusses the Reconnaissance, Resource Development, and Initial Access tactics.

The Initial Access tactic is the most frequently employed, typically used to establish an adversarial foothold in the target network \cite{varlioglu2024pulse, park2022performance, mustafa2024threat}. Prominent sub-techniques include T1078 (Valid Accounts) and T1566 (Phishing), with T1078 gaining attention in 2024. Several studies emphasize these sub-techniques. \cite{abdullah2024effective} examines T1566.001 (Spearphishing Attachment) and T1566.002 (Spearphishing Link) for enhancing adversary detection and improving SOC capabilities. \cite{lanni2024boosting} identifies T1566 as a frequent tactic in high-severity attacks against governmental and industrial entities in Asia. \cite{rahman2022investigating} reveals a strong correlation between T1566 and other techniques, such as T1204 (User Execution), with a predictive likelihood of 0.95, and underscores its frequent co-occurrence with techniques like T1105 (Ingress Tool Transfer) and T1059 (Command and Scripting Interpreter) in multi-step attack scenarios. \cite{ccakmakcci2023apt} applies T1566.001 and T1078 in an attack simulation to detect anomalous activities in Microsoft applications.

Reconnaissance remains a key element in adversarial campaigns \cite{landauer2022maintainable, bagui2022detecting, yi2024anomaly}. For instance, \cite{bagui2022detecting} highlights the use of tree-based models for detecting reconnaissance activities through network connection features. \cite{yi2024anomaly} analyzes reconnaissance in APT campaigns, using communication port fluctuations and small data packets to capture characteristics like network penetration and information gathering. Graph databases and star motifs \cite{bagui2023introducing} provide innovative ways to model reconnaissance. T1595 (Active Scanning) is notably prominent in this tactic. \cite{landauer2022maintainable} identifies a vulnerable wpDiscuz plugin during the Reconnaissance phase using Nmap and WordPress scanning. Studies like \cite{mayukha2022reconnaissance} and \cite{shin2023beyond} demonstrate the role of T1595 in information gathering and planning attacks, such as DoS.

The Resource Development tactic has received limited attention but remains crucial for adversarial preparation \cite{belfadel2022towards, liao2021generating, youn2022research}. T1588 (Obtain Capabilities) is the most observed technique. \cite{batalla2024threat} uses this tactic to mitigate military network threats through structured attack path analysis. \cite{al2023analysis} highlights T1588 and T1583 (Acquire Infrastructure) among the top 20 techniques in Campaign threat profiles, though it remains underutilized in pre-attack phases. In operational contexts, \cite{belfadel2022towards} applies T1588 to identify system vulnerabilities in autonomous vehicle systems. \cite{skjotskift2024automated} identifies T1588.002 (Tool Acquisition) as a frequently observed technique in acquisition of malicious tools for advancing attacks.

\subsection{Mid Attack-Stage Focus} 

Mid attack stages usually involve Execution, Persistence, Privilege Escalation , Defense Evasion, Credential Access, Discovery, Lateral Movement tactics.

The Execution tactic, the third most frequently observed, involves adversaries executing malicious code to achieve their objectives \cite{sharma2023ttp, lewis2024effects, perez2022evaluation}. Techniques such as T1059 (Command and Scripting Interpreter) and T1204 (User Execution) are commonly used. \cite{arreche2024xai, arreche2024two} highlight T1059 as a frequent technique in network intrusions, while \cite{song2023similarity} identifies it as a predominant technique in the Execution phase of Hive ransomware attacks. In another study, \cite{perez2022evaluation} contrasts initial user-triggered techniques like T1204 with system-based techniques like T1059, which exploit built-in OS functionalities to bypass detection \cite{rahman2022investigation}. \cite{katano2022prediction} investigates the combined use of T1204.002 (Malicious File) and T1059.001 (PowerShell) in analyzing device activity logs.

The Persistence tactic enables adversaries to maintain a foothold in compromised systems despite interruptions \cite{lee2022icstasy, takey2022real, park2023performance}. Common techniques include T1547 (Boot or Logon Autostart Execution) and T1543 (Create or Modify System Process), which ensure that malware or scripts restart with the system. T1137 (Office Application Startup) embeds malicious macros in Office documents to execute automatically. \cite{perez2022evaluation} demonstrates how attackers use T1137 to maintain persistence through embedded malicious macros. However, \cite{virkud2024does} notes inconsistencies in the application of T1547 across detection rules, highlighting the need for standardization in persistence technique usage.

The Privilege Escalation tactic involves exploiting vulnerabilities to gain elevated access to systems \cite{maynard2020big, chimmanee2024digital, amro2023cyber}. Techniques such as T1543 (Create or Modify System Process), T1078 (Valid Accounts), and T1055 (Process Injection) are commonly used. For example, \cite{maynard2020big} reveals several techniques employed by hacktivist Phineas Fisher to gain elevated privileges within compromised systems, such as monitoring user activity, exploiting vulnerabilities, capturing user input, hijacking authenticated sessions, and modifying services to steal credentials.

The Defense Evasion tactic allows adversaries to bypass security measures and avoid detection \cite{ajmal2021offensive, gylling2021mapping, amro2022click}. Techniques such as T1055 (Process Injection), T1036 (Masquerading), and T1562 (Impair Defenses) are used to obfuscate malicious actions. \cite{amro2022click} describes methods to avoid detection on the victim ship's network, such as using encoded command and control messages. \cite{mccarty2023cybersecurity} explores T1055, where arbitrary code is injected into a running process’s address space, enabling unauthorized access and privilege escalation. On the defense side, \cite{perez2022evaluation} discusses using Event-Based Detection Systems (EBDS) to detect attacks involving T1055 and T1219 (Indirect Command Execution).

The Credential Access tactic targets user or administrative credentials to gain unauthorized access to systems or accounts \cite{maynard2020big, varlioglu2024pulse, nisioti2021game}. Common techniques include T1003 (OS Credential Dumping) and T1110 (Brute Force), while techniques such as T1555 (Credentials from Password Stores) and T1212 (Exploitation for Credential Access) are less frequently reported but remain critically relevant. \cite{alsabbagh2024investigating} highlights T1110.002 (Password Cracking), which exploits intercepted network traffic to decrypt passwords for unauthorized access. \cite{varlioglu2024pulse} discusses how Brute Force attacks and Mimikatz are used in fileless cryptojacking scenarios.

The Discovery tactic is the second most frequently observed and involves adversaries gathering information about systems to prepare for further attack stages \cite{sharma2023ttp, gulatas2023malware, lee2023game}. Techniques such as T1082 (System Information Discovery) and T1083 (File and Directory Discovery) are most prominent, showing an upward trend from 2022 to 2024. \cite{gonzalez2023technical} reports that T1082 is more common in malware campaigns (9.23\%) than in APT attacks, as APT groups conduct extensive reconnaissance, while malware dynamically collects system information during execution. \cite{sharma2023ttp} also discusses T1124 (System Time Discovery), which utilizes network flows on port 123 used by NTP for system time discovery.

The Lateral Movement tactic enables adversaries to move within a network to access additional systems or data \cite{gourisetti2022assessing, song2023similarity, ccakmakcci2023apt}. Techniques such as T1021 (Remote Services) exploit remote access protocols (e.g., RDP, SSH, and SMB) to control other systems. \cite{ccakmakcci2023apt} uses T1021 to detect malicious lateral movement and identify anomalies in network activity. \cite{song2023similarity} identifies a common set of techniques, including lateral tool transfer and remote service utilization, employed by various ransomware families for lateral movement.

\subsection{Late Attack-Stage} 

Late attack stages mainly refer to Collection, Exfiltration, Command and Control and Impact tactics. 

The Collection tactic enables adversaries to gather critical data from compromised systems for further operations or exfiltration \cite{chimmanee2024digital, batalla2024threat, lee2023game}. Common techniques include T1056 (Input Capture) and T1005 (Data from Local System). \cite{afenu2024industrial} utilizes the Collection tactic to analyze Modbus/TCP network traffic and detect ARP cache poisoning attacks. \cite{gonzalez2023technical} reports that T1005 is prevalent in APT attacks, appearing in 26.59\% of cases, as it helps attackers locate sensitive files for exfiltration. Similarly, \cite{youn2022research, shin2022focusing} highlight the use of T1005 and T1056 by the Kimsuky APT group, which exploits phishing attacks to steal local system data. \cite{okada2024predicting} maps log data to T1056 and other techniques.

The Command and Control (C2) tactic establishes covert communication between adversaries and compromised systems \cite{amro2022click, yi2024anomaly, lee2023game}. Techniques like T1071 (Application Layer Protocol) and T1105 (Ingress Tool Transfer) are frequently observed. \cite{li2019extraction} identifies T1071 as one of the top 10 most commonly used threat actions, enabling covert communication through protocols like HTTPS and DNS. \cite{amro2022click} uses AIS as a covert channel to send commands and transfer files, with AIS message type 8 being particularly suitable due to its large capacity and flexibility.

The Exfiltration tactic involves extracting data from compromised systems while evading detection \cite{lee2023game, Mohamed2022air, shen2024decoding}. Techniques like T1041 (Exfiltration Over C2 Channel) highlight the importance of secure communication surveillance, while less common techniques like T1052 (Exfiltration Over Physical Medium) provide additional insights. \cite{Mohamed2022air} establishes T1052.001 (Exfiltration Over USB) as a method for exfiltrating data from air-gapped networks, using tools like the Bash Bunny to circumvent privilege escalation constraints. \cite{lee2023game} analyzes data exfiltration from the perspectives of Medium (transmission protocols like HTTP/S and tools like rclone.exe) and Enclosure (methods to enhance stealth, such as limiting transfer packet size).

Impair Process Control tactic has received limited attention but targets the disruption of control processes in OT environments \cite{mumrez2023comparative, lee2023assessment, ekisa2024leveraging}. Key techniques include T0829 (Manipulate I/O Image) and T0830 (Change Program State), which exploit vulnerabilities in PLCs and HMIs to alter system behavior. \cite{mumrez2023comparative} provides insights into the vulnerability of smart grid infrastructures, showing how parameter modifications and unauthorized command injections could risk operational continuity, leading to availability loss and forced shutdowns. 

The Inhibit Response Function tactic, which focuses on disabling or tampering with safety mechanisms, occurs infrequently in the reviewed studies \cite{gourisetti2022assessing, liao2021generating, hotellier2024standard}. Key techniques include T0896 (Disable Safety Instrumented System) and T0897 (Tamper with Safety Mechanisms). \cite{liao2021generating} applies this tactic in an ICS attack scenario, using T0881 (Service Stop) to halt critical processes such as plant operations, and suggests monitoring network traffic and application logs to detect endpoint DoS attacks.

The Impact tactic aims to manipulate, interrupt, or destroy systems and data, causing operational or reputational damage \cite{pell2021towards, lasky2023machine, hotellier2024standard}. It is observed across Enterprise, Mobile, and ICS frameworks. T1499 (Endpoint Denial of Service) is the most prevalent technique, while T1490 (Inhibit System Recovery) and T1491 (Defacement) are less common but carry significant disruptive potential. \cite{francik2023connecting} identifies T1499.004 (Network Denial of Service) as one of the top three techniques with critical risks in municipal infrastructure.

\section{Validation Methods} \label{sec:Validation}

This section discusses how various studies validated the effectiveness of ATT\&CK in enhancing security practices.

\subsection{Case Studies} 

Case studies provide in-depth analyses of real-world or simulated scenarios to validate the practical applicability of MITRE ATT\&CK integrated methods. A recurring theme is the use of real-world data for validation. For example, \cite{villanueva2023analyzing} presents three case studies analyzing vulnerabilities in GE Proficy, Siemens SIMATIC WinCC, and Schneider Electric Triconex Tricon systems. These studies demonstrate how CWE data, ICS ATT\&CK information, and CISA advisories can help understand attack patterns and identify techniques exploiting specific weaknesses. Similarly, \cite{dev2023models} uses 207 incidents from the VERIS Community Database (VCDB) to validate its framework for categorizing privacy breaches by threat actor, mechanism, and impact. \cite{nisioti2021data} evaluates the DISCLOSE framework using three realistic cyber incident scenarios, comparing it to a baseline and CBR-FT framework variations with metrics such as the percentage of revealed actions and relevance of recommended inspections.

Case studies often span diverse sectors, illustrating multi-domain applicability. For instance, \cite{masi2023securing} applies a cybersecurity digital twin methodology to transport and road tunnel systems to simulate attacks and evaluate countermeasures, while \cite{oruc2022assessing} assesses maritime systems by evaluating cyber risks in real-world integrated navigation systems. These studies highlight the versatility of MITRE ATT\&CK-based frameworks in addressing sector-specific security needs.

Some studies validate frameworks through real-world use case scenarios. For example, \cite{sadlek2022identification} demonstrates the KCAG model’s application in a multi-step attack scenario targeting a personal computer, illustrating how the model represents attack paths and categorizes steps into kill chain phases. \cite{mavroeidis2021threat} evaluates a threat actor inference methodology using the Lazarus Group, demonstrating its efficacy with cases like the DarkSeoul attack and Sony Pictures Entertainment breach.

Several studies focus on categorizing and describing incidents comprehensively. \cite{dev2023models} assesses its framework’s ability to categorize privacy breaches based on specific elements like threat actors and mechanisms. \cite{choi2021probabilistic} tailors Hidden Markov Model (HMM) parameters to emulate malware behaviors such as Triton, validating the framework with realistic attack sequences. \cite{choi2021probabilistic} also analyzes real-world ICS incidents and malware reports like Stuxnet, BlackEnergy, and Industroyer, verifying attack sequences through HMMs. These simulations test the adaptive and predictive capabilities of the systems. Additionally, some studies focus on evaluating vulnerabilities or risk metrics, such as \cite{yoon2023vulnerability}, which applies the Exploit Risk Score (ERS) methodology to the Juniper Networks Junos OS J-Web vulnerability (CVE-2023-36844).

\subsection{Empirical Experiment} 

Experimental validation involves testing MITRE ATT\&CK integrated frameworks, models, or methods in controlled environments to assess their performance against predefined metrics or criteria. Two primary approaches are used: ML-based model validation and attack scenario generation or assessment validation.

In ML-based validation, models are evaluated across diverse datasets to assess their ability to detect, classify, and predict cyber threats. Common metrics, such as accuracy, recall, precision, and F1 score, are used to measure performance. For example, \cite{branescu2024automated} explores transformer-based models (e.g., SecBERT, SecRoBERTa, CyBERT, TARS, T5) for CVE-to-tactic mapping, using a weighted F1 score to evaluate models on a dataset of 9985 CVEs. \cite{yi2024anomaly} assesses various classification models with 10-fold cross-validation for both binary and multi-class tasks. \cite{sen2023approach} trains MLP and LSTM models on synthetic IDS alert datasets, evaluating performance using accuracy and recall metrics against synthetic data and a lab-simulated WannaCry attack.

The evaluation of specific model components is also a focus. \cite{huang2024mitretrieval} evaluates the MITREtrieval framework using a diverse CTI report dataset, focusing on minimizing false negatives using the F2 score and analyzing components like the topic classifier and knowledge fusion process. Visualizations are used to enhance understanding. Similarly, \cite{toure2024framework} employs a hybrid approach combining supervised and unsupervised learning, using F1 scores for supervised models and Silhouette scores to evaluate clustering quality in K-Means, validating potential zero-day attacks through online learning with both industrial (IBM) and public (NSL-KDD) datasets.

Attack scenario generation or assessment validation simulates real-world cyber threat environments to test predictive MITRE ATT\&CK integrated mechanisms. For example, \cite{nour2024automa} evaluates the AUTOMA framework’s ability to generate relevant attack hypotheses, focusing on hypothesis ranking, reduction rates, and execution efficiency. Using a dataset of 86 APTs and 350 campaigns, the study demonstrates AUTOMA's ability to prune irrelevant hypotheses and quickly identify relevant threats, with APT41 used to showcase performance under varying conditions. Similarly, \cite{ccakmakcci2023apt} constructs a controlled environment using the ELK stack and adversary emulation techniques to measure the correlation engine’s ability to detect APTs through alert generation. \cite{mukherjee2024proviot} uses realistic datasets, including benign data from 33 ARM-based IoT devices and malicious data from IoT malware and simulated APT scenarios aligned with MITRE ATT\&CK.

Real-world environments are also employed for validation. \cite{brenner2023better} tests a safety-augmented NIDS in a pilot factory environment, simulating attacks such as port scans, botnet creation, and DoS. The network traffic is analyzed to train a random forest classifier. \cite{hotellier2024standard} validates an IDS method using both experiments and case studies, assessing scalability and detection accuracy on a Schneider Electric testbed with specific attack scripts. \cite{toker2021mitre} describes a multi-level evaluation for an EtherCAT-based water management system, focusing on both the ICS network level (PLCs) and the endpoint systems level (engineering workstations). Real-world datasets also play a crucial role in validation. \cite{havlena2022accurate} combines experimental testing with benchmark datasets containing IEC 104 and MMS communication protocols from real and simulated ICS environments to assess an anomaly detection method.

\subsection{Simulation}

Simulations replicate cyberattack scenarios in virtual environments to assess the resilience and response of systems or frameworks. Studies frequently use diverse virtual testbeds and frameworks to model and evaluate attack scenarios. For example, \cite{skjotskift2024automated} applies Markov Chain Monte Carlo (MCMC) simulations, where states represent sets of techniques and abilities. By sampling the state space based on transition probabilities, the study estimates the probability distribution of attack chains, providing insights into adversarial behavior. Similarly, \cite{alfageh2023water} simulates a water desalination plant in Saudi Arabia using MATLAB Simulink to test 20 attack scenarios, assessing the impact on operational parameters and the effectiveness of Multilevel Bayesian Networks (MBNs) and Dynamic Programming (DP) methods for cyberattack identification and mitigation. \cite{ekisa2024leveraging} uses the MITRE ATT\&CK framework to evaluate simulated attack scenarios within the VICSORT testbed, highlighting its utility in structuring and assessing adversarial behaviors. In another study, \cite{afenu2024industrial} designs a virtual ICS environment to execute DoS and ARP poisoning attacks, monitoring network traffic and system behavior to validate security tools like Suricata and Snort integrated with MITRE ATT\&CK.

\subsection{Qualitative Assessment} 

Qualitative assessments evaluate methods based on subjective criteria such as expert opinions, usability, or interpretability. These assessments often involve interviews, surveys, or questionnaires to evaluate the applicability of MITRE ATT\&CK-integrated approaches in real-world scenarios or their alignment with established standards \cite{kern2024logging}. For example, \cite{aghamohammadpour2023architecting} uses a survey of cybersecurity experts to assess twelve quality attributes through pairwise comparisons, gathering qualitative judgments on their relative importance. Similarly, \cite{derbyshire2021talking} combines interviews with ten cyber risk practitioners and offensive security professionals to validate their framework’s feasibility and practicality, supported by a scenario-based exercise to identify cost factors in real-world attacks. Grounded theory is applied in \cite{mcneil2020analysis} through interviews with fourteen adversarial cyber testing professionals, using open, axial, and selective coding to uncover patterns and insights. \cite{omiya2019secu} evaluates the Secu-One framework using two qualitative evaluation models (ARCS and MEEGA+), supplemented by participant questionnaires and free-form feedback. In \cite{franklin2017toward}, semi-structured interviews are used to validate the approach, with the first round focusing on understanding analysts' workflows, tools, and challenges, and the second round mapping network activity data to the MITRE ATT\&CK framework. These interviews in \cite{franklin2017toward} provide critical insights during the development stage, contributing significantly to validation.

\section{Discussion}
\label{sec:Discussion}

This section provides an overview of the current state of knowledge, highlighting the framework's strengths, challenges, and potential future directions in cybersecurity research and practice.

\subsection{Primary Uses of MITRE ATT\&CK and Effectiveness}

The MITRE ATT\&CK framework is widely applied across various cybersecurity domains and industries. Key research areas, including Cyber Threat Intelligence (30.2\%, or 126 out of 417 papers), Threat Hunting (17.5\%), Threat Modeling (10.8\%), and Attack Simulation (10.1\%), underscore the framework’s broad utility (Section \ref{sec:FocusAreas}). It demonstrates adaptability across industries such as IT (28.8\%, or 120 out of 417 papers), Manufacturing (14.4\%), and Communications (6.7\%) systems, with 43.1\% of research falling under the General category, highlighting its broad applicability. However, sectors like Energy (2.2\%), Transportation (2.4\%), Media (1.2\%), and Healthcare (1.2\%) remain underrepresented, indicating a need for more focused applications (Section \ref{sec:ApplicationDomain}).

ATT\&CK is crucial for structuring threats by mapping adversarial TTPs to data sources such as vulnerability reports and IDS events, enabling multi-stage attack tracing (Section \ref{sec:TTPs}). This capability is especially useful in improving detection precision and recall in identifying APTs and ransomware \cite{sen2023approach, al2023grid, yameogo2024improving}. This framework also aids in evaluating the effectiveness of EDR and ML-based detection tools by mapping them to adversarial tactics. ATT\&CK supports threat hunting by modeling adversarial behavior and improving threat visibility through graph-based models and chronological sequencing, allowing proactive identification and mitigation of threats \cite{fujita2023structured, bagui2024graphical}. Additionally, it enhances attack simulations by providing realistic environments to test system defenses against APTs and other attacks \cite{skjotskift2024automated, bierwirth2024design}. In incident response, ATT\&CK maps adversarial actions to guide mitigation strategies, improving response times and minimizing the impact of breaches \cite{chimmanee2024digital, al2024threat}. As a training tool, it enhances professionals' understanding of adversary tactics, improving their ability to respond effectively to complex scenarios \cite{bierwirth2024design, nisioti2021game}.

The reviewed studies employ a range of tools and techniques to gather and analyze data, including CTI platforms, vulnerability assessment tools, and monitoring systems like MISP \cite{mundt2024enhancing}, Splunk \cite{virkud2024does}, Wazuh \cite{ammi2023cyber}, and VMware Carbon Black \cite{patil2023audit} (Section \ref{sec:Framework}). These tools facilitate the aggregation of threat intelligence and the monitoring of malicious activities, while simulation tools like Atomic Red Team \cite{orbinato2024laccolith} and Caldera \cite{chetwyn2024modelling} replicate adversarial tactics to validate detection methods. Data from diverse sources, such as NVD \cite{grigorescu2022cve2att}, ExploitDB \cite{ampel2023mapping}, APT reports \cite{ajmal2023toward}, and domain-specific logs like Syslog \cite{mustafa2024threat} and Zeek \cite{alkhpor2023collaborative}, are mapped to the MITRE ATT\&CK framework for identifying attack patterns (Section \ref{sec:DataSources}). ML and NLP techniques, including transformer-based models like BERT \cite{alves2022leveraging} and traditional algorithms like Random Forest \cite{alkhpor2023collaborative} and SVM \cite{ampel2023mapping}, are applied to analyze datasets and map them to adversarial tactics (Section \ref{sec:Analytical}). Additionally, ATT\&CK is integrated with other cybersecurity frameworks, such as \cite{chimmanee2024digital} and STRIDE \cite{davis2024cyber}, and supported by tools like ATT\&CK Navigator \cite{van2023mitre} for visualizing and analyzing attack behaviors.

The effectiveness of using the MITRE ATT\&CK framework in cybersecurity research is often validated using a combination of quantitative and qualitative methods (Section \ref{sec:Validation}). More specifically, empirical experiments dominate the evaluation landscape, and are commonly used to test security tools, algorithms, and frameworks in controlled environments with datasets or simulated attacks aligned with MITRE ATT\&CK \cite{branescu2024automated, nour2024automa}. Metrics such as detection accuracy, false positive rates, and time to detection are used to evaluate system performance \cite{sen2023approach, yi2024anomaly}. Case studies map adversarial behaviors to ATT\&CK techniques, assessing detection and mitigation capabilities with metrics tailored to specific use cases, such as revealed attack actions and resource consumption \cite{dev2023models, choi2021probabilistic}. Simulations replicate real-world attack scenarios in virtual environments to evaluate system resilience and security controls, with metrics varying by study objectives \cite{skjotskift2024automated, alfageh2023water}. These methods validate the framework’s effectiveness in dynamic environments. Qualitative assessments, though less frequent, use expert feedback from interviews and surveys to evaluate the practical applicability of ATT\&CK-based approaches, often using scoring systems or comparisons against established standards \cite{aghamohammadpour2023architecting, franklin2017toward}.

\subsection{Challenges of MITRE ATT\&CK Application}

The MITRE ATT\&CK framework has seen significant application across cybersecurity domains, but several challenges remain in its implementation. The most studied tactics are Initial Access, Execution, and Discovery, which together account for over 30\% of mentions, highlighting the focus on early attack-stage detection and mitigation (Section \ref{sec:TTPs}). In contrast, tactics such as Resource Development, Impair Process Control, and Inhibit Response Function are underexplored but gaining attention, particularly in ICS-related research.

Despite its widespread adoption, MITRE ATT\&CK faces several key challenges. As cyber threats evolve rapidly, maintaining accurate and current mappings of TTPs requires significant resources and expertise \cite{yameogo2024improving}. Studies highlight the need for robust update mechanisms to ensure the framework's relevance and accuracy, addressing issues of outdated or incorrect mappings \cite{gabrys2024using, kim2024relative}. In addition, high-level abstractions in ATT\&CK often lack the granularity needed for specific domains such as ICS and IoT \cite{hotellier2024standard, gjerstad2022lademu}. Domain-specific adaptations are necessary, yet many tools lack built-in mappings, requiring manual tagging or custom scripts, which complicates integration and limits practicality \cite{georgiadou2021assessing}.
    
Mapping real-world behaviors to ATT\&CK techniques is a resource-intensive and subjective process, often prone to bias \cite{kim2023design, hobert2023enhancing}. Semantic gaps between ATT\&CK techniques and other data sources, such as API descriptions, further complicate mapping \cite{sajid2021soda}. Effective mapping relies heavily on dataset quality and expert involvement, limiting scalability and generalizability \cite{villanueva2023analyzing}. Meanwhile, organizations with limited resources, such as small security teams or insufficient computational power, face significant barriers in adopting ATT\&CK \cite{zhang2024vtt, boakye2023implementation}. Mapping large volumes of network traffic or logs to the framework is computationally intensive, leading to delays and bottlenecks, especially in less mature SOCs \cite{abdullah2024effective}. Also, ATT\&CK’s abstract concepts often do not align with real-world data, such as discrepancies between honeypot data and real-world attack scenarios \cite{gjerstad2022lademu}. Incomplete datasets or regulatory restrictions on data sharing can further hinder the mapping of the full attack lifecycle to ATT\&CK techniques \cite{park2022performance}. Privacy concerns complicate the collection and processing of data for ATT\&CK-based analyses \cite{dev2023models}.

\subsection{Future Directions}

To address these challenges, continuous updates and the development of automated TTP mapping tools are critical. Real-time detection and response systems, integrated with SIEM platforms and SOC tools, could reduce manual intervention and streamline processes. Future research may explore the use of graph neural networks, deep learning, and hybrid AI approaches to enhance scalability and the accuracy of cybersecurity solutions.

A prominent future direction involves expanding datasets to include a wider range of real-world attack scenarios. Integrating diverse data sources, such as system logs and network traffic, will enhance threat analysis' granularity. Addressing underrepresented TTPs, particularly in IoT and ICS, is also a key focus.

Expanding the scope of MITRE ATT\&CK to cover additional tactics, sub-techniques, and specialized domains—such as maritime systems, healthcare networks, and 5G infrastructures—remains a significant challenge. Improved integration with existing tools and scalable solutions for processing large datasets will enhance usability. Additionally, developing privacy-preserving methods for data analysis will ensure more comprehensive and secure threat analysis.

Validation and evaluation efforts continue to underpin future research. Digital twins, synthetic datasets, and temporal-spatial analyses are being employed to replicate complex environments, enabling thorough evaluations of ATT\&CK-integrated defense strategies, particularly in IT/OT networks. Expanding these simulations to critical infrastructures will address sector-specific cybersecurity gaps. Ongoing work will also focus on reducing false positives and refining detection mechanisms to enhance reliability.

\section{Conclusion}
\label{sec:Conclusion}

The MITRE ATT\&CK framework has become foundational in advancing cybersecurity practices by enabling precise threat identification, categorization, and proactive mitigation. Its structured TTPs have driven significant progress, particularly in Cyber Threat Intelligence (CTI) extraction, threat hunting, and modeling. Through the application of NLP and ML techniques, ATT\&CK has facilitated the automation of TTP extraction from unstructured CTI reports, enhancing the efficiency and understanding of adversarial behaviors.

A systematic review of 417 articles demonstrates the widespread integration of MITRE ATT\&CK with established cybersecurity methodologies such as the Cyber Kill Chain, NIST guidelines, STRIDE, and CIS Controls. The reviewed studies leverage diverse data sources, including system and network logs, vulnerability databases, and CTI reports, as well as tools like Wireshark, Elastic Stack, VirusTotal, and MISP. This integration strengthens threat detection, adversarial modeling, and overall defense against complex attacks.

Despite its success, MITRE ATT\&CK faces significant challenges, particularly in maintaining relevance due to the rapid evolution of cyber threats, the complexity of mapping TTPs to real-world scenarios, and the resource-intensive nature of updates. Additionally, the framework's need for domain-specific adaptations and the computational burden of comprehensive mappings hinder its broader application.

Future research should focus on automating CTI extraction and TTP mapping, utilizing advanced ML techniques such as graph neural networks and large language models (LLMs) to enhance threat detection and response. Further improvements could involve enriching ATT\&CK with more detailed contextual information on threat actors and their motivations, enabling more targeted defenses. Expanding ATT\&CK to address emerging technologies (e.g., cloud computing, IoT, blockchain) and sector-specific threats (e.g., healthcare, finance, critical infrastructure) is also essential for adapting to evolving cybersecurity challenges.

\bibliographystyle{ACM-Reference-Format}
\bibliography{MITRE}


\end{document}